\documentclass[usenatbib]{mnras}

\usepackage{newtxtext,newtxmath}

\usepackage{tikz}
\usetikzlibrary{arrows.meta}

\usepackage[T1]{fontenc}

\DeclareRobustCommand{\VAN}[3]{#2}
\let\VANthebibliography\thebibliography
\def\thebibliography{\DeclareRobustCommand{\VAN}[3]{##3}\VANthebibliography}
\usepackage{xcolor}
\usepackage{caption}

\usepackage{graphicx}	
\usepackage[thinc]{esdiff}
\usepackage{braket}

\usepackage{amsmath}	
\usepackage{mathtools}
\usepackage{hyperref}
\hypersetup{colorlinks=true,
            citecolor=blue,
            linkcolor=blue}
\usepackage[export]{adjustbox}

\usepackage{chngcntr}
\usepackage{newtxtext,newtxmath}

\newcommand{\msun}{\mathrm{M}_{\odot}}	
\newcommand{\rh}{r_{\rm H,3}}
\newcommand{\rhb}{r_{\rm H,12}}
\newcommand{\rht}{r_{\rm H,t}}

\newcommand{\Et}{E_{\rm trip}}

\newcommand{\asin}{a_\mathrm{sin}}
\newcommand{\abin}{a_\mathrm{bin}}

\newcommand{\pbin}{r_\mathrm{p,bin}}

\newcommand{\mSMBH}{M_\bullet}

\defcitealias{Rowan2022}{Paper~I}
\defcitealias{Rowan2023}{Paper~II}
\begin{document}


\title[Hydrodynamical Binary-Single encs]{Prompt gravitational-wave mergers aided by gas in Active Galactic Nuclei:
The hydrodynamics of binary-single black hole scatterings
}
\author[C. Rowan et al.]{
Connar Rowan$^{1,2}$\thanks{E-mail: connar.rowan@nbi.ku.dk}, Henry Whitehead$^{3}$, Gaia Fabj$^{1}$, Pankaj Saini$^{1}$, Bence Kocsis$^{2,4}$, \newauthor $\,\,$Martin Pessah$^{1}$ and Johan Samsing$^{1}$
\\
$^{1}$Niels Bohr International Academy, Niels Bohr Institute, Blegdamsvej 17, DK-2100 Copenhagen Ø, Denmark 
\\
$^{2}$Rudolf Peierls Centre for Theoretical Physics, Clarendon Laboratory, University of Oxford, Parks Road, Oxford, OX1 3PU, UK
\\
$^{3}$Department of Physics, Astrophysics, University of Oxford, Denys Wilkinson Building, Keble Road, Oxford OX1 3RH, UK
\\
$^{4}$St Hugh's College, St Margaret's Rd, Oxford, OX2 6LE, UK
\\
}


\date{\today}


\maketitle

\begin{abstract}
 Black hole binary systems embedded in AGN discs have been proposed as a source of the observed gravitational waves (GWs) from LIGO-Virgo-KAGRA. Studies have indicated binary-single encounters could be common place within this population, yet we lack a comprehensive understanding of how the ambient gas affects the dynamics of these three-body encounters. We present the first hydrodynamical simulations of black hole binary-single encounters in an AGN disc. We find gas is a non-negligible component of binary-single interactions, leading to unique dynamics, including the formation of quasi-stable hierarchical triples. The gas efficiently and reliably dissipates the energy of the three-body system, hardening the triple provided it remains bound after the initial encounter. The hardening timescale is shorter for higher ambient gas densities. Formed triple systems can be hardened reliably by $2-3$ orders of magnitude relative to the initial binary semi-major axis within less than a few AGN orbits, limited only by our resolution. We calculate that the gas hardening of the triple enhances the probability for a merger by a minimum factor of $3.5-8$ depending on our assumptions. In several cases, two of the black holes can execute periapses on the order of less than $10$ Schwarzschild radii, where the dynamics were fully resolved for previous close approaches. The likelihood of these prompt mergers increases when the gas density is larger. Our results suggest that current timescale estimates (without gas drag) for binary-single induced mergers are an upper bound. The shrinkage of the triple by gas has the prospect of increasing the chance for unique GW phenomena such as residual eccentricity, dephasing from a third object and double GW mergers.
\end{abstract}



\section{Introduction}
\label{sec:intro}
The astrophysical mechanisms driving compact object mergers and observed gravitational wave (GW) sources \cite{LIGO2016,LIGO2019,LIGO2020a,LIGO2020b,LIGO2020e,LIGO2020c,LIGO2020d,abbott2022,abbott2022_fermi,KAGRA:2021duu, KAGRA:2021vkt} remain an open question. Possible environments which would allow black holes (BHs) to reach small enough separations to merge within a Hubble time include isolated stellar binary evolution \citep[e.g.][]{Lipunov1997,Belczynski2010,Belczynski2016,Dominik2012,Dominik2013,Dominik2015,Tagawa2018,Giacobbo2018}, dynamical three-body or even four-body interactions in globular clusters \citep[e.g.][]{Zwart2000,Mouri2002,Miller2002,Downing2010,Rodriguez+2015,Rodriguez+2016,Rodriguez+2018,Fragione_Kocsis2019,Samsing2018,Zevin+2019,DiCarlo2020,Hamers2020,Liu2021,Arca_Sedda+2021,Antonini2023,Samsing2024,Hendriks2024} and in AGN \citep[e.g][]{OLeary+2009,Bartos2017,Stone2017,Tagawa2018,Hoang+2018,Yang2019,Tagawa2020, Tagawa2020_spin,Tagawa+2021_hierarchical,Tagawa+2021_eccentricity,Grobner2020,McKernan2020_AGNmontecarlo,Ford2022,Rom2024,Vaccaro2024,Wang2024,Delfavero2024,Rowan2024_rates}. 

AGN present a unique environment due to the presence of a highly dense and thin accretion disc, in addition to a large spherical stellar component, i.e the nuclear stellar cluster (NSC). These two components play host to many different dynamical processes. In the NSC, we have two-body, binary-single and binary-binary scatterings \citep[e.g][]{Antonini+2012,Antonini+2016, Trani2024}. Objects within the NSC that cross the AGN disc can become embedded through a combination of accretion/gas dynamical drag \citep[e.g][]{Bartos2017,Panamarev2018,Yang2019,Fabj2020,Wang2024} and vector resonant relaxation \citep{Rauch1996,Kocsis2015,Szolgyen2018,Gruzinov2020,Magnan+2022,Mathe+2022}. The disc-like geometry of this population, compared to a spherical component, may increase the frequency of the aforementioned BH interactions due to the enhanced number density of objects and through disc migration \citep[e.g.][]{Secunda2019,Secunda2021}. In addition to the same kind of encounters within the NSC, the high gas density of the disc facilitates the efficient formation of black hole binaries (BBHs) through two-body scatterings by dissipating the relative two-body energy of the BHs through gas drag, known as ``gas-captured'' binaries \citep{Tagawa2020}. This BBH hydrodynamical formation mechanism has been studied through hydrodynamical simulations of the global disc \citep[e.g.][]{Rowan2022,Rowan2023}, shearing box simulations \citep[e.g.][]{Li_Dempsey_Lai+2022,Whitehead2023,Whitehead2023_novae} and N-body simulations using analytical gas drag treatments \citep[e.g.][]{DeLaurentiis2022,Rozner+2022,Dodici2024}. 

In this paper we examine the influence of gas on binary-single interactions which are expected to be common during the evolution of disc-embedded binaries if the typical gas-driven inspiral timescale is greater than the collisional timescale for 3-body interactions (as suggested by \citealt{Tagawa2020}). Binary-single interactions play a dominant role in shaping the merging mass distribution and the spin-orbit angle distribution due to exchange interactions and by reorientating the binaries' orbital plane. The timescale for the inspiral of embedded BBHs is still highly uncertain, as the spatial scales involved to accurately model the hydrodynamics from the BBH's formation to merger can be over seven orders of magnitude, with estimates between $10^{2}-10^{6}$ year \citep[e.g][]{Baruteau2011,Vaccaro2024,Ishibashi2024,Dittman2024_analytic}. There is a growing amount of evidence that retrograde binaries (orbiting in the opposite direction with respect to the orbit around the AGN) experience eccentricity pumping, which may accelerate the merger timescale \citep[e.g][]{Dittman2022, Rowan2022,Calcino2024}. The expected merger and eccentricity pumping timescale scales inversely with the density of the AGN disc \citep[e.g][]{Baruteau2011,Ishibashi2020,Li_and_lai_2022,Rowan2022}. Therefore, given the immense range in expected AGN disc densities across the various initial parameters in AGN disc models, this can lead to highly varying timescales regardless of the method used to estimate the binary torques. %

The dynamics of three body scatterings are inherently chaotic \citep[e.g][]{Aarseth1974,Heggie1974,Hut1983,Samsing2018_topology}, yet three (or more) body systems are abundant throughout the Universe. Examples include asteroids \citep{Kozai1962,Lidov1962}, planetary binaries \citep{Nagasawa2008,PeretsNaoz2009}, stellar triples  \citep{Eggleton2001,Duchene2013,Toonen2016}, stellar compact object triples \citep{Thompson2011}, black hole binaries in AGN \citep[e.g][]{Leigh2018,Trani2019,Ginat2021,Samsing+2022,Fabj2024,Trani2024,Rom2024} and SMBH systems \citep{Blaes2002}. 

The stability of a three body system depends on the relative separations between the objects. In a hierarchical triple, one object exists at a large separation from a more tightly bound binary system. Provided the separation is sufficiently large, the outer object may only weakly perturb the inner binary orbit and vice versa over secular timescales \citep[e.g][]{Ford2000,Toonen2016}. If there is sufficient inclination between the inner and outer binary orbit, then secular oscillations in the eccentricity and inclination driven by the Kozai-Lidov \citep{Kozai1962,Lidov1962} oscillations could alter the periapsis of the inner binary, speeding up the timescale for merger in the case of BHs \citep[e.g][]{Naoz+2013,Antonini2014,Naoz2016,silsbee2017_triple}. If the separations are comparable, the evolution is non-secular and the objects continually exchange energy and angular momentum, until one object is ejected from the system (in the absence of any dissipation mechanism). \cite{Fabj2024} frame this case as a sequence of binary-single like states, where the eccentricities and semi-major axes may be sampled from probability distributions each time the less bound third object executes its periapsis. The continual ``re-shuffling'' of the binary eccentricity during the evolution of non-hierarchical BH systems has profound implications for the expected eccentricity distributions of GW signals, where we expect binaries may be able to merge with residual eccentricity through dynamical BH triple interactions in AGN \citep[e.g][]{Ishibashi2020} and globular clusters \citep[e.g][]{Samsing2014,Dallamico2024}. Constraining the rates of these highly eccentric mergers from both theory and GW observations will shed light on the proportion of BH mergers coming from dynamical channels vs isolated binary evolution.

Although analytic and numerical N-body studies present exciting prospects for unique observational GW features and possibly shorter BH merger timescales in AGN, they currently neglect the AGN disc component. As indicated by the efficiency of the gas-capture mechanism during single-single interactions, the gas may also significantly affect the dynamics during a binary-single encounter in the AGN disc. In this work we perform the first  simulations of binary-single encounters in a gaseous accretion disc that include the effects of the gas, using a fully hydrodynamical approach. We examine how gaseous drag alters the dynamics of the initial encounter as well as the evolution of formed black hole triple systems. In later sections, we calculate the GW merger probabilities and their enhancement when gas is included to quantify the significance of  gaseous effects. We describe the simulations and their initial conditions in Sec. \ref{sec:methods}, before presenting our results in Sec. \ref{sec:results}. The implications of our results are discussed in Sec. \ref{sec:discussion}. We summarise and conclude our results in Sec. \ref{sec:conclusions}. 
\section{Methods}
\label{sec:methods}  
This work extends our previous studies of single-single black hole encounters \citep{Rowan2022,Rowan2023,Whitehead2023_novae,Whitehead2023} in AGN to the binary-single case. We simulate the encounters and the hydrodynamics using the Eulerian GRMHD code \texttt{Athena++} \citep{Stone_2020}. We utilise a second-order accurate van Leer predictor-corrector integrator with a piecewise linear method (PLM) spatial reconstruction and Roe's linearised Riemann solver. See \citet{Stone_2020} for  a detailed discussion of the integration scheme used in \texttt{Athena++}. We simulate a 2D portion of the AGN disc in a rectangular shearing box \citep[e.g][]{Goldreich1978,Hawley1994,Hawley1995}.
The important length scales for the system are the single, binary and 3-body Hill spheres: $r_{\rm H,i}, r_{\rm H,b}$ and  $\rht$ respectively:
\begin{align}
    r_{\rm H,i} =&\,R_0\bigg(\frac{M_{\rm i}}{3\mSMBH}\bigg)^{1/3}\label{rh1},\\
    r_{\rm H,b}=&\,R_0\bigg(\frac{M_\mathrm{bin}}{3\mSMBH}\bigg)^{1/3}\label{rh2},\\
    r_{\rm H,t}=&\,R_0\bigg(\frac{M_\text{trip}}{3\mSMBH}\bigg)^{1/3}\label{rh3},        
\end{align}
where $M_{\rm bin} = M_1+M_2$ is the mass of the binary, $M_{\rm trip} = M_\text{bin}+M_3$ is the mass of the triple, $\mSMBH$ is the supermassive black hole  (SMBH) mass and $R_0$ is the radial position of the shearing box from the SMBH. We define $i=(1,2)$ as the BHs present in the initial (pre-encounter) binary and $i=3$ as the incoming single BH. 
\subsection{Nomenclature}
Concerning the language of triple encounters, use of the term \textit{close triple encounter} will specifically refer to an encounter where the separations of the three objects are \textit{comparable} and within $\rht$, during the three-body interaction (which may occur multiple times). During a triple encounter, the terms \textit{single} or \textit{BH} pertain to the currently least bound object to the 3-body centre of mass (COM). The terms \textit{binary} or \textit{BBH} refer to the currently most energetically bound pair of BHs. The semi-major axes, eccentricities and periapses for the binary and single components during the three body scattering are denoted $a_\mathrm{bin}$, $e_\mathrm{bin}$ and $r_\mathrm{p,bin}$ for the binary and $a_\mathrm{sin}$, $e_\mathrm{sin}$ and $r_\mathrm{p,sin}$ respectively. The binary quantities are calculated assuming purely 2-body dynamics and the single quantities are similarly calculated assuming the binary is a point mass with mass $M_i+M_j$ and $i,j\in\{1,2,3\}$.
\subsection{The shearing box}
We utilise a comparable shearing box setup to \cite{Whitehead2023_novae,Whitehead2023}, where the encounter takes place in a non-inertial reference frame that co-rotates with the AGN disc at a fixed radius $R_0$ and angular frequency $\Omega_0=\sqrt{G\mSMBH/R_0^{3}}$. The Cartesian coordinate system of the shearing box $\{x,y\}$ can be translated to a position in the global AGN disc in cylindrical coordinates $\{R,\phi\}$ as
\begin{equation}
    \boldsymbol{r}=\bigg(\begin{matrix}
        R \\ \phi
    \end{matrix}\bigg)
    =\bigg(\begin{matrix}
         R_0 + x  \\
        \Omega_0 t+y/R_0 
    \end{matrix}\bigg)\,,
\end{equation}
The gas and BHs are subject to accelerations from the Coriolis and centripetal forces. These can be added together as 
\begin{equation}
    \label{eq:a_smbh}
    \boldsymbol{a}_{\text{SMBH}}= 2 \boldsymbol{u} \times \Omega_0 \boldsymbol{\hat{z}} + 2q\Omega_0^2 (x-x_\mathrm{C})\hat{x}\,,
\end{equation}
where $\boldsymbol{u}$ 
represents here the gas or BH velocity in the corotating frame, $x_\mathrm{C}$ is the $x$ position of co-rotation and $q=-\frac{d\ln\Omega}{d\ln R}=\frac{3}{2}$ is the shear parameter for a Keplerian disk. Thus, the equilibrium trajectories where Eq. \eqref{eq:a_smbh} is zero are 
\begin{equation}
    \boldsymbol{u}_\mathrm{eq}=\bigg(\begin{matrix}
        u_{\rm x} \\ u_{\rm y}
    \end{matrix}\bigg)
    =\bigg(\begin{matrix}
         0  \\
        -q\Omega_0 (x-x_\mathrm{C})
    \end{matrix}\bigg)\,.
\end{equation}
These trajectories represent Keplerian circular orbits of varying radii in the global picture, but straight lines in the tangential $\boldsymbol{\hat{y}}$ direction in the shearing frame.

The shearing box has a radial width of $0.24R_{0}$ and azimuthal extent $0.96R_{0}$\footnote{Note that the global dynamics of the disc may not be as accurately captured here compared to a global disc setup as the azimuthal extent of the disc is large. However we must maintain this azimuthal extent in order to reach steady state in the minidiscs prior to encounter. We expect that any changes in the dynamics will arise from deviations in the gas mass within $\rht$ during the triple encounter. As such, this is unlikely to significantly affect our findings as this will only affect the dissipation timescale, see later in Sec \ref{sec:param_space}.}, 
corresponding to $\sim18\rht$ and $\sim72\rht$ respectively (see Sec. \ref{sec:BHs}). Defining $x_{\rm C}$ as the $x$ position of the shear in the box about the midpoint $(x=0)$, the boundary conditions at the upper ($y=y_\mathrm{max}$) and lower edge ($y=y_\mathrm{min}$) of the box are:
\begin{equation}
    (y=y_\mathrm{max}):\,\,\begin{cases}
        \text{outflow} & x<x_\mathrm{C},\\
        \text{refill} & x>x_\mathrm{C},\\        
    \end{cases}
\end{equation}
\begin{equation}
    (y=y_\mathrm{min}):\,\,\begin{cases}
        \text{refill} & x<x_\mathrm{C},\\
        \text{outflow} & x>x_\mathrm{C}.\\        
    \end{cases}
\end{equation}
The boundary conditions at the $x$ boundaries are set to outflow. The refill regions assume the initial gas velocities and densities at the start of the simulation, i.e $\{u_\mathrm{x},u_\mathrm{y}\}=\{0,-q\Omega_0 x\}$ and $\Sigma=\Sigma_{0}$ (see Sec. \ref{sec:AGNdisc}).  
\subsection{Gas dynamics}
\label{sec:gas_dynamics}
\texttt{Athena++} solves the fluid equations in Eulerian form through the extended Navier-Stokes equations, in 2D: 
\begin{align}\label{eq:EOM}
    &\frac{\partial \Sigma}{\partial t} + \nabla \cdot \left(\Sigma \boldsymbol{u}\right) = 0\,,\\
    &\frac{\partial \left(\Sigma \boldsymbol{u}\right)}{\partial t} + \nabla \cdot \left(\Sigma \boldsymbol{u} \boldsymbol{u} + P \boldsymbol{I} + \boldsymbol{\Pi}\right) = \Sigma\left(\boldsymbol{a}_{\text{SMBH}} - \nabla \phi_\text{BH}\right)\,.\label{eq:EOM2}
\end{align}
Here, $\Sigma$, $\boldsymbol{u}$, $P$, $\Pi$, are the cell gas surface density, velocity, pressure and viscous stress tensor
\begin{equation}\label{eq:stress}
    \Pi_{ij} = \Sigma \nu \left(\frac{\partial u_i}{\partial x_j} + \frac{\partial u_j}{\partial x_i} - \frac{2}{3}\delta_{ij}\nabla \cdot \boldsymbol{u}\right)\,,
\end{equation}
for a viscosity $\nu$. The viscosity is modelled using the standard $\alpha$-disc approach, where $\nu=\alpha c_{\rm s}H$ and $H=c_{\rm s}/\Omega$ is the disc scale height. In all simulations here, $\alpha=0.1$. We assume an isothermal equation of state where $P=\Sigma c_{\rm s}^{2}$ and the sound speed $c_{\rm s}$ is fixed. We ignore any magnetic effects and the self-gravity of the gas.

The remaining $\nabla\phi_{\rm BH}$ term is the acceleration from the potential of all ($n_\mathrm{BH}=3$) stellar mass BHs,
\begin{equation}
    -\nabla \phi_\text{BH}(\boldsymbol{r}) = \sum^{n_\text{BH}}_{i=1} M_i \,g\left(\frac{\boldsymbol{r}-\boldsymbol{r}_n}{h}\right)\,, 
\end{equation}
where $g(s)$ is the gas gravitational softening kernel, see \cite{Price2007}
\begin{equation}
    g(\boldsymbol{s}) = -\frac{G}{h^2}\hat{\boldsymbol{s}}
    \begin{cases}
    \frac{32}{3}s - \frac{192}{5}s^3 + 32s^4 & 0 < s \le \frac{1}{2} \\
    -\frac{1}{15s^2} + \frac{64}{3}s - 48s^2 + \frac{192}{5}s^3 - \frac{32}{3}s^4 & \frac{1}{2} < s \le 1 \\
    \frac{1}{s^2} & s > 1\,.
    \end{cases}
    \label{eq:softening}
\end{equation}
As 3-body encounters can often lead to chaotic trajectories, we cannot rule out very close BH-BH separations. Therefore, we set the softening length $h$ to be smaller than our previous work at $h_{i}=0.005r_{\rm H,i}$. This allows us to trust the dynamics of these close encounters as the enclosed mass within $r<h$ (where gas gravitation is not accurately described) is very low and unlikely to alter the trajectories of the BHs significantly should they execute trajectories within $h$. Note that only the gas-BH interactions are softened, but the BH-BH interactions are not (see \S~\ref{sec:BHs}).
\subsection{Mesh Refinement}
We apply an adaptive mesh refinement (AMR) procedure, allowing us to resolve the area around the BHs to a high degree, whilst minimising compute time and resources. The mesh closer to the location of each BH becomes more refined, down to a minimum refinement level. In all simulations shown here, we maintain a base resolution of 256x1024, with 8 mesh refinement levels. For our box size, this gives a maximal and minimal cell size of $\delta_\mathrm{max}\simeq0.1\rht$ and $\delta_\mathrm{min}\simeq 0.00039\rht$, The softening length is then resolved by $\rht/\delta_\mathrm{min}\gtrsim10$ cells across one dimension. The AMR scheme is centred on the positions of the BHs and moves with them throughout the simulation, ensuring the necessary resolution around the BHs at all times, see \cite{Whitehead2023} for more detail.
\subsection{Initial conditions}
\subsubsection{The AGN disc}
\label{sec:AGNdisc}
To model an accurate AGN disc, we determine the ambient surface density $\Sigma_{0}$ and sound speed $c_{\rm s,0}$ of the gas in the shearing box from AGN disc profiles generated using \texttt{pAGN} \citep{Gangardt2024}. In this paper, we consider a fiducial setup assuming an AGN disc with an Eddington fraction $L_\mathrm{\epsilon}=0.1$, radiative efficiency $\epsilon=0.1$ and hydrogen/helium fractions X/Y = 0.7/0.3. We consider an AGN with a SMBH mass at the peak of the anticipated merger rate distribution for gas-captured binaries predicted in \cite{Rowan2024_rates}, where $\mSMBH=10^{7}\msun$. The shearing box radius is set to $R_0=0.01$pc. For these parameters, this gives $\Sigma_0\simeq1.6\times10^{5}$kg$\,$m$^{-2}$, $c_{\rm s,0}\simeq11.9$km$\,$s$^{-1}$ and disc thickness ratio $H/R\simeq0.0057$. We investigate how the AGN disc density affects the triple dynamics by considering two additional values such that we have three disc densities $\Sigma=\{0.1,0.25,1\}\Sigma_0$, where we take 
$\Sigma=\Sigma_0$ to be our fiducial value.
\subsubsection{The black holes}
\label{sec:BHs}
For simplicity, we consider three representative equal mass BHs ($M_1=M_2=M_3=25\msun$). This gives the mass scales for a single BH and triple system: $q_1=M_i/\mSMBH=2.5\times10^{-6}$ and  $q_3 = \sum_\mathrm{i=1}^{3}M_i/\mSMBH =7.5\times10^{-6}$, respectively. The BHs are represented by non-accreting point-like  particles. The motion of these BHs are determined purely by gravity through direct summation of the forces between the BH sinks, background SMBH frame forces (Eq. \ref{eq:a_smbh}) and the gas in each cell. Like the gas, the gravitational accelerations due to cells within $h$ of a BH is softened according to Eq. \eqref{eq:softening}. The BH-BH interactions remain unsoftened. The gravitation of the BHs is purely Newtonian and hence dissipation from GWs is not considered explicitly, though we discuss its implication in Sec. \ref{sec:mergers}.
\subsubsection{The encounter}
The binary and single system are initialised with zero inclination and marginal eccentricity of $e_0=0.02$, as a result all scatterings will also be co-planar. In this 2D geometry the scattering is parametrised by the radial impact parameter $p=x_\mathrm{s}-x_\mathrm{b}$, where $x_\mathrm{s}$ and $x_\mathrm{b}$ are the $x$ positions of the single and the binary COM respectively at the moment the single object is introduced. The initial binary semi-major axis $a_0$ is set to $0.1r_\mathrm{H,b}$. We prioritise standardising the semi-major axis of the binary at the encounter and therefore turn off the gravitation of the BHs due to the gas until they reach a separation of $2\rht$. Not doing so would introduce variable hardening rates on the binary for the different initial gas densities considered here, leading to a correlation between the assumed ambient density and the initial conditions (binary semi-major axis) of the encounter, complicating the analysis. 

In our previous shearing box simulations of single-single BH scatterings \citep{Whitehead2023,Whitehead2023_novae}, the time required for the gas to form a steady morphology around each object was short enough to permit their construction naturally within a reasonable azimuthal extent of the shearing box. As we are now dealing with a binary, the time required for the gas around the binary to form a steady state is longer. Therefore, we initialise only the binary in the simulation for $\sim 50$ binary orbits before inserting the single BH, to allow the formation of a well defined and steady circum-binary discs. After this settling time, we initialise the single with impact parameter $p$ and an azimuthal displacement that is normalised such that the approach time for each encounter is the same as that for $p=1.5\rht$ and an initial azimuthal distance $\Delta y=30\rht$, i.e $\Delta y/\rht =30 p/1.5\rht$ . These initial conditions allow the binary $\sim 120$ orbits to fill its Hill sphere and reach a steady morphology before the encounter with the single.

We perform simulations with 24 different impact parameters in the range $p=[1.2,1.9]\rht$ for each value of the density $\Sigma$ for a total of 72 simulations. An initial linear sweep over the full range was performed with 15 simulations. We then doubled the sampling uniformly in the parameter range which leads to chaotic triple encounters ($p=[1.375,1.775]\rht$, see Sec. \ref{sec:param_space}) with a further 9 simulations.
\subsubsection{Shifting the shearing frame}
At the beginning of each simulation, we anticipate the location of the scattering using the 3-body COM. We initialise the initial binary's position such that the 3-body COM lies in the origin when the single is inserted. In this work, we consider the scattering of a binary on an inner orbit compared to the single, therefore the binary is located in a position towards the lower left of the simulation domain. To allow the gas around the binary to form a steady morphology before inserting the single, we fix the co-rotating position $x_\mathrm{C}$ initially on the binary (i.e $x_\mathrm{C,0}=-2p/3$) so it does not have any motion in the shearing box. After five AGN orbits, we insert the single and shift $x_\mathrm{C}$ to the origin and boost the velocity of all sinks and gas cells to match the new co-rotation frame with the velocity boost given by $\Delta v_\mathrm{y}=q\Omega_0 x_\mathrm{C,0}=-\frac{2}{3}q\Omega_0 p$.
\begin{figure}
    \centering
    \includegraphics[width=8.5cm]{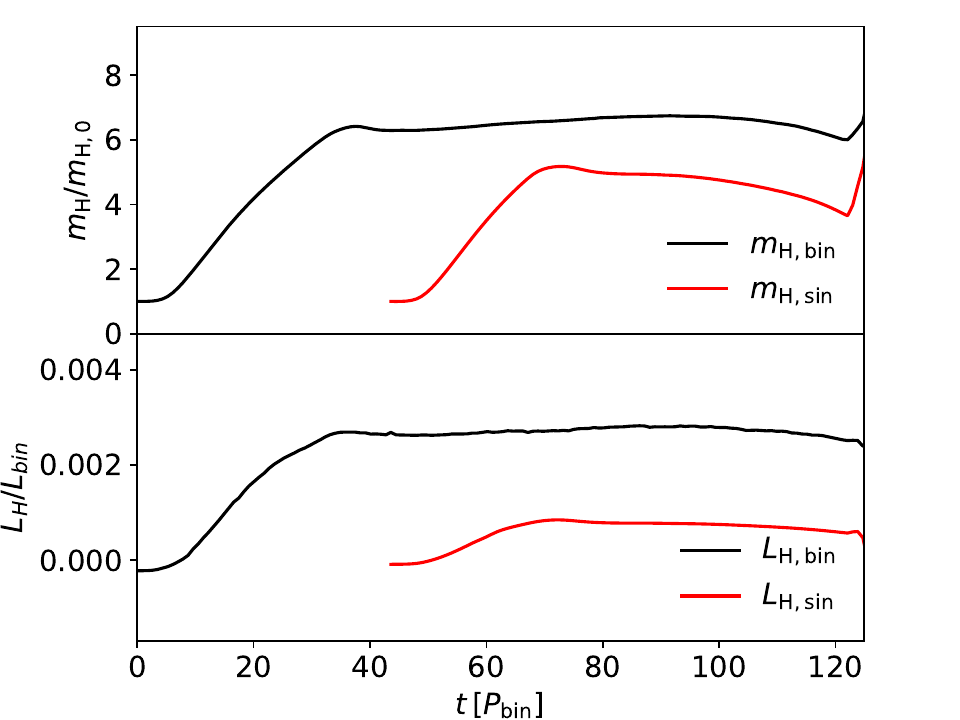}
    \caption{
    Time evolution of gas properties in the vicinity of the binary and single prior to encounter for the simulation with $p=1.7\rh$ and $\Sigma=0.1\Sigma_0$. \textit{Top:} the gas mass $m_\mathrm{H}$ contained within the Hill sphere of the binary ($r<\rhb$) and single ($r<\rh$), normalised to the mass initially enclosed within their respective Hill radii $m_\mathrm{H,0}$. \textit{Bottom:} the angular momentum of the enclosed gas $L_\mathrm{H}$ (as measured from the single and the binary COM) normalised to the angular momentum of the binary $L_\mathrm{bin}$ . Both quantities are shown as a function of time in inner binary orbits $P_\mathrm{orb}$. The binary Hill mass reaches a steady state after $50 P_\mathrm{orb}$ and the single after $40 P_\mathrm{orb}$, respectively.}
    \label{fig:hillmass}
\end{figure}
\begin{figure*}
    \centering
    \includegraphics[width=17cm]{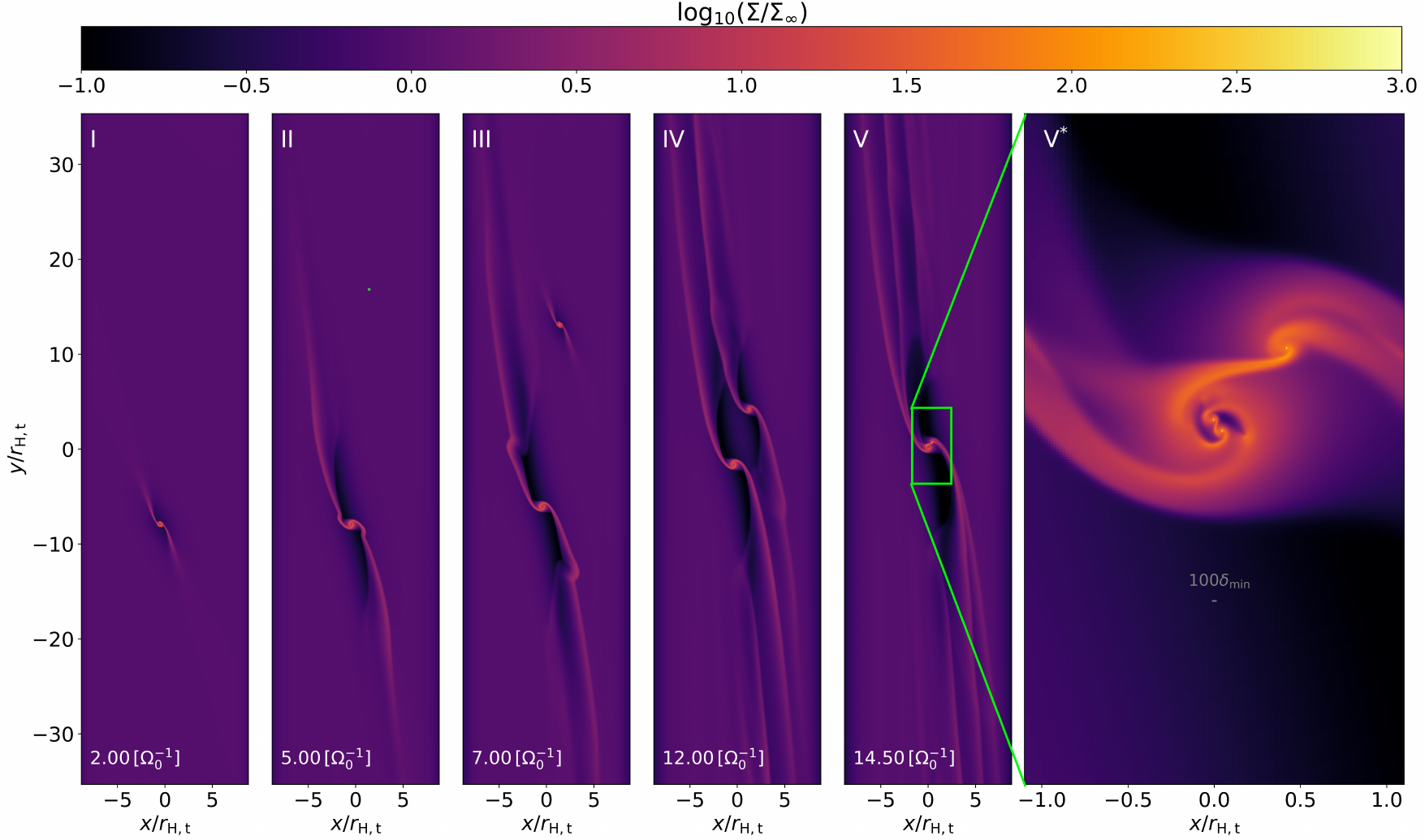}
    \caption{The timeline of a binary-single scattering simulation ($\Sigma=0.1\Sigma_0$, $p=1.4\rht$), visualised through the surface density $\Sigma$ normalised to the initial ambient value $\Sigma_\mathrm{\infty}$. \textit{Panel I}: The binary is initialised in its offset position with the shear centred on its COM. \textit{II:} The single is injected (green dot) and the radial location where the shear vanishes $x_\mathrm{C}$ is shifted to the origin, which is the COM of the three-body system. \textit{III:} The single forms a circum-single disc as both the single and binary approach the encounter location at the origin. \textit{IV:} With both components now in steady state, they enter the under-dense gas region just prior to encounter (see Sec. \ref{sec:steady_state}). \textit{V:} The system undergoes a close encounter. \textit{V*:} A zoom in of the close encounter as the colliding gas forms a shock. The small grey line indicates the length scale of 100 times the smallest cell size $\delta_\mathrm{min}$.}
    \label{fig:timeline}
\end{figure*}
\subsubsection{Reaching steady state}
\label{sec:steady_state}
The most influential parameter for the dissipation during embedded object encounters is the mass within each object's Hill sphere $m_\mathrm{H}$. If the encounter occurs before the mass has time to build up and reach a steady value, the energy/angular momentum exchange between the objects and the gas will typically be lower. In Figure \ref{fig:hillmass} we show the gas mass $m_\mathrm{H}$ and angular momentum $L_\mathrm{H}$ enclosed within the Hill sphere of both the binary and the single prior to the encounter. 
The binary Hill mass and angular momentum reaches a steady state after approximately 50 inner binary orbits ($P_\mathrm{orb}$), remaining approximately flat until the encounter at $120P_\mathrm{orb}$. The gas within the Hill sphere of the single reaches a steady state after roughly $40P_\mathrm{orb}$. As the single nears the binary ($\sim100P_\mathrm{orb}$), the gas flow into their respective Hill spheres changes due to interaction with the undersense regions associated with the gas morphology around each perturber, leading to a reduction in $m_\mathrm{H}$ prior to encounter. The approach time for the encounter was set to the minimum that allows $m_\mathrm{H}$ in the single to reach a steady state before this reduction takes place.

A visual timeline of one of our simulations is shown in Figure \ref{fig:timeline} through visualisations of the surface density $\Sigma$ in the full extent of the shearing box. All density visualisations in this work are normalised to the initial ambient value $\Sigma_{\infty}$. In the Figure, we show the initial fixed position of the binary as it begins to form a circum-binary disc, followed by the insertion of the single, the approach of the binary and single, the encounter and a close-up of the encounter. We also compare the resolution to the scale of the binary, demonstrating the binary is well resolved by $\sim 250$ cells over its semi-major axis.

\section{Results}
\label{sec:results}
\subsection{Fiducial example}
\begin{figure*}
    \centering
    \includegraphics[width=16cm]{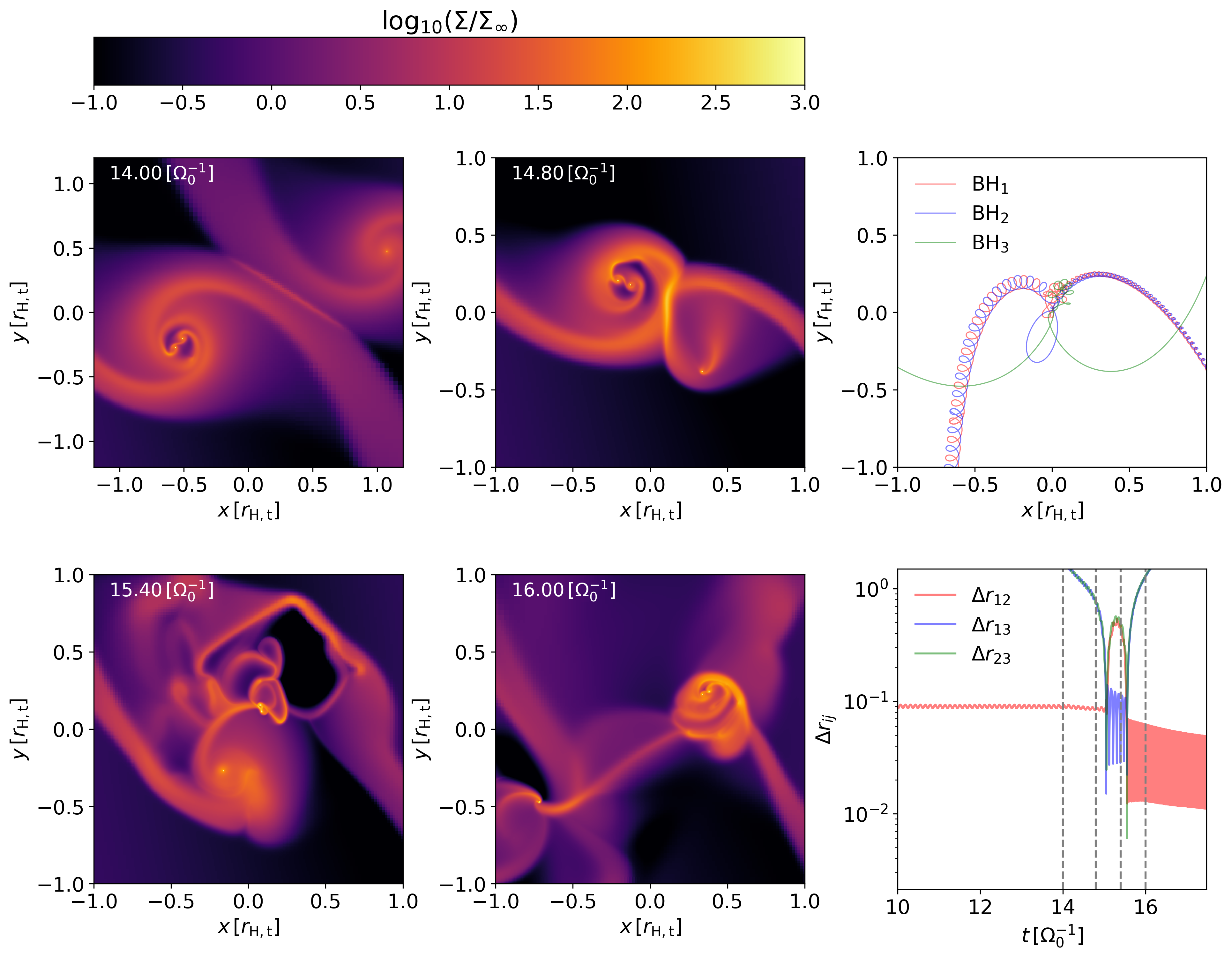}
    \caption{An example encounter with $\Sigma=0.1\Sigma_0$ and $p=1.7\rht$. \textit{Top left:} the gas morphology at the moment the minidiscs of the binary and single collide, the colours represent the surface density relative to the ambient value of the simulation $\Sigma_{\infty}$. \textit{Top centre:} the morphology of the gas as the the single begins plunging into the binary. \textit{Bottom left:} The gas morphology following the first major encounter of the third object. Here, one binary object has been replaced by the single via an exchange interaction. \textit{Bottom centre:} The gas morphology after one object has been removed from the system, hardening the remaining binary. \textit{Top right:} The trajectories of the three BHs in the 3-body COM. \textit{Bottom right:} The separation between each BH $\Delta r_{ij}$ as a function of time. The vertical dashed lines represent the timestamps of the $\Sigma$ plots.}
    \label{fig:example}
\end{figure*}

As a qualitative example, we show the trajectories and separations of the three BHs in addition to the gas morphology at four intervals from one of our simulations ($p=1.7$, $\Sigma=0.1\Sigma_0$) in Figure \ref{fig:example}.
The first panel shows the initial shock formed when the gas in the binary and single Hill spheres come into contact. Rapidly after this point, both components lose their initially well-defined discs and the morphology becomes highly complex and variable. In this example, the interaction leads to a binary swap ($t\sim15.1\Omega_0^{-1}$), visible in the separation between each of the three BHs. The bottom left panel shows the formation of the temporary binary system formed when the single swaps place with one of the other BHs. This is then reversed around $t=15.6\Omega_0^{-1}$, where another close triple encounter leads to the re-pairing of the original binary BHs and the third is ejected, leaving the binary hardened compared to both the original and the temporary binary formed after the first encounter. While the three-body dynamics play out, the gas leads to a net hardening of the temporary binary and also continues to harden the final binary after the third is ejected. This fiducial example demonstrates \textit{simultaneous} hardening of a BH through a three-body exchange and gas drag.
\subsection{Dissipation}
The closest analogue to the binary-single encounters in gas is the single-single gas-capture mechanism. During such an encounter, the two-body energy of the BHs is dissipated to the gas through a combination of accretion (direct linear momentum transfer) or gas dynamical drag (dynamical linear momentum transfer). The amount of energy that can be transferred is approximately proportional to the gas contained within the Hill sphere of each approaching object \citep[e.g][]{Rowan2023,Whitehead2023} as this effectively serves as the available reservoir to deposit the energy of the BHs. For a three body system, the energy of BHs is given by the three body energy
\begin{equation}
    \centering
    E_{\rm trip} = \sum_{i}^{3}\frac{1}{2}M_i v_i^{2}-\sum_{j}^{3}\sum_{i>j}^{3}\frac{G M_i M_j}{r_{i,j}}\,,
    \label{eq:E3body}
\end{equation}
where $v_i$ is the velocity of BH $i$ relative to the 3-body COM and $r_{i,j}=\|\boldsymbol{r}_i-\boldsymbol{r}_j\|$ is the separation of BHs $i$ and $j$.
In the two body case with BHs $i$, $j$ the energy is given by 
\begin{equation}
    \centering
    E_{\rm bin} = \frac{1}{2}\mu \|v_i-v_j\|^{2} - \frac{G M_i M_j}{r_{i,j}}\,,
    \label{eq:E2body}
\end{equation}
where $\mu=M_1 M_2/M_\text{bin}$ is the reduced mass of the binary.

\begin{figure}
    \centering
    \includegraphics[width=7cm]{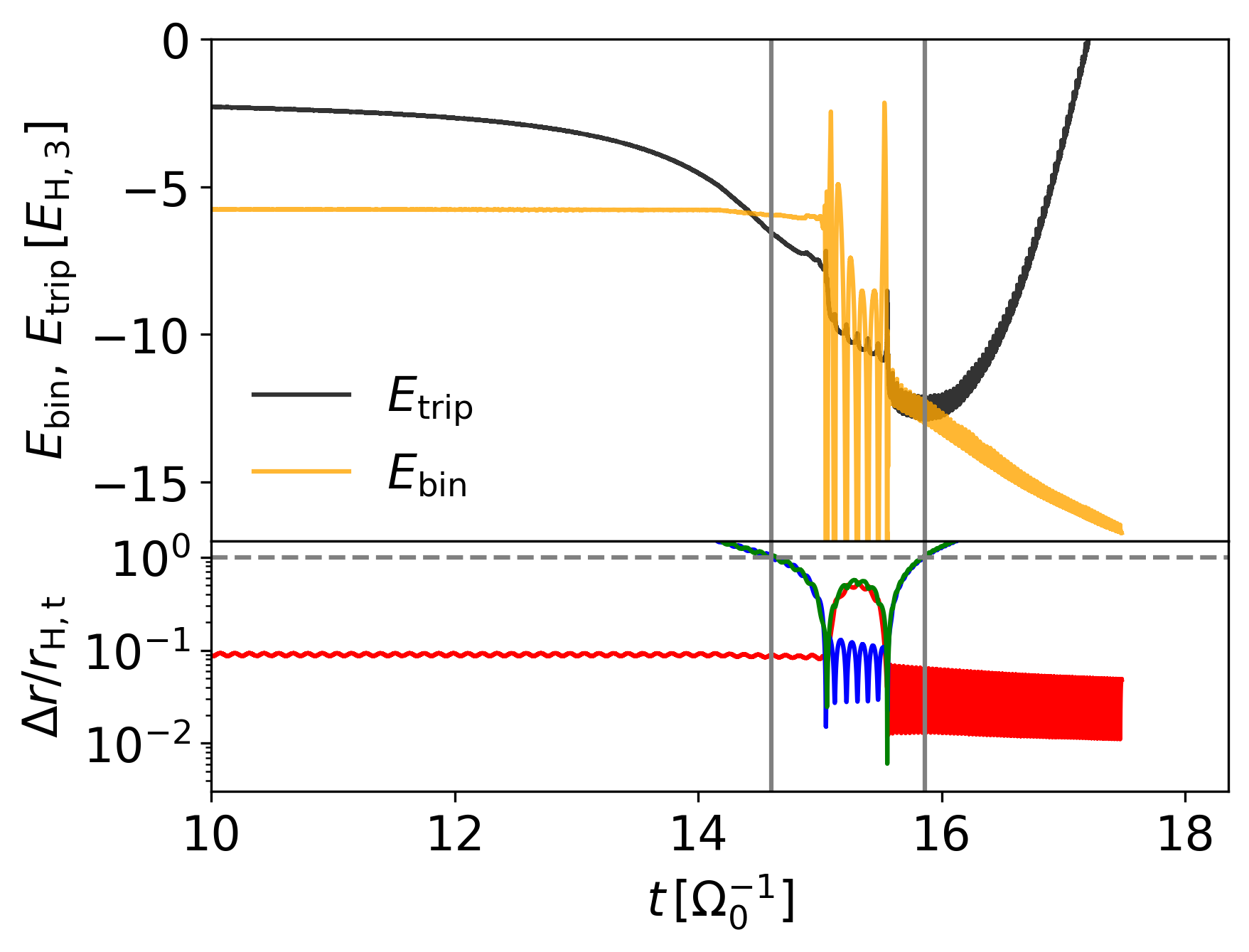}
    \caption{The evolution of the three-body energy $\Et$ in the fiducial simulation. \textit{Top:} The three body energy of the system (Eq. \ref{eq:E3body}) and energy of the most bound BH pair (Eq. \ref{eq:E2body}) as a function of time in units of $\Omega_{0}^{-1}$. \textit{Bottom:} The separation between each of the three objects $\Delta r_{ij}$. The vertical grey lines border the time period where all three objects are within $r_\mathrm{H,t}$.}
    \label{fig:dissipation}
\end{figure}

We show how $E_\mathrm{trip}$ and $E_\text{bin}$ for the most bound pair of BHs changes throughout the fiducial model in Figure \ref{fig:dissipation}.
We define a natural energy scale for the system $E_\mathrm{H,3}$ as the absolute energy of a two body system with masses $M_3$ and $M_1+M_2$, and a separation of $r_\text{H,t}$, i.e.
\begin{equation}
    \centering
    E_\mathrm{H,3}= \frac{G(M_1+M_2)M_3}{2r_\mathrm{H,t}}
    \label{eq:EH3}
\end{equation}
At the first encounter, a large amount of energy is dissipated from the system by the gas. This initial large energy loss is consistent with single-single scatterings, owed to the large gas mass contained in the Hill sphere of the binary and single that they accumulated before the encounter. Following the first encounter and the binary object exchange, the newly formed triple system is hardened slightly (visible in the more gradual decline in $E_\mathrm{trip}$) before another close 3-body interaction occurs and one BH is ejected. At the second close encounter, there is another spike in the dissipation. We attribute this to the single BH re-accumulating gas while executing its large apoapsis before the second close encounter. The remaining binary is hardened (observe the second gradual decline in $E_\mathrm{bin}$ and $E_\mathrm{trip}$ following the second encounter). When the ejected BH leaves $r_\mathrm{H,t}$ the SMBH potential cannot be ignored and $E_\mathrm{trip}$ becomes unreliable as a metric for the system's hardness.
\subsection{Alternate encounter outcomes}
\label{sec:encounter_types}
\begin{figure*}
    \centering
    \includegraphics[width=17cm]{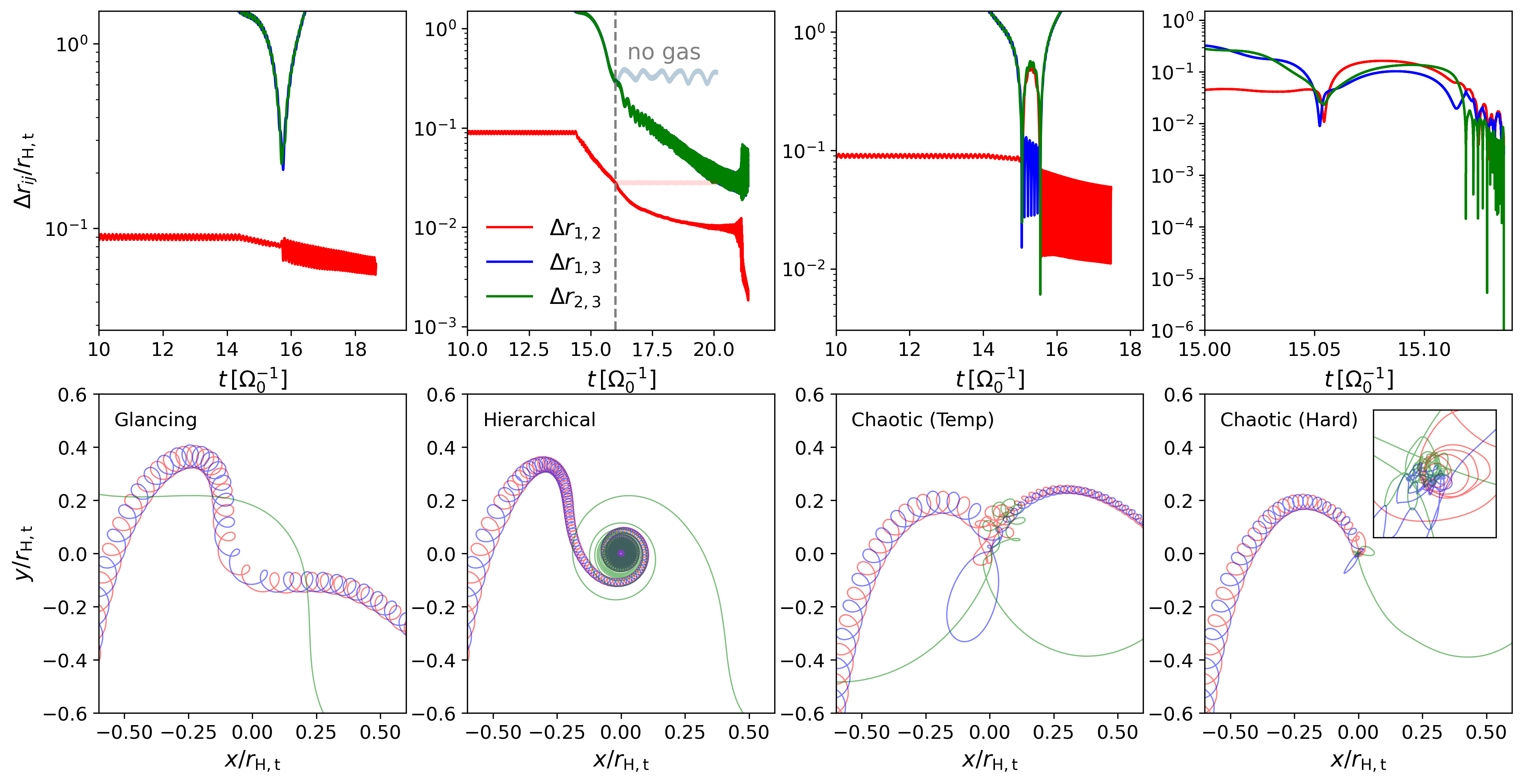}
    \caption{The trajectories and object separations for the four characteristic types of triple encouners in gas. \textit{Left to right:} Glancing encounter, hierarchical encounter, temporary chaotic encounter, hardened chaotic encounter. \textit{Top row:} The separation between each pair of BHs $\Delta r_{ij}$ as a function of time. \textit{Top row:} the trajectories in $x$ and $y$ of each BH during the encounter in the 3-body COM frame. In the hierarchical panel, we turn off gas effects at $t=16\Omega^{-1}$ (vertical dashed line), observing the stagnation of the single BHs semi-major axis.}
    \label{fig:encounter_types}
\end{figure*}

The dynamics of the encounters and their outcomes can be divided into characteristic families. 
\begin{itemize}
    \item \textit{Glancing} encounters - where the minimum separation of the third body remains on the order of or larger than $\rht$ and only a weak interaction takes place.
    \item \textit{Hierarchical} encounters - when a single approaches and is successfully captured through gas into a hierarchical system where the single orbits the original binary and evolves in a secular fashion.
    \item \textit{Temporary chaotic} encounters - where the periapsis of the single passes close to the binary and gas dissipation removes an insufficient amount of energy, leading to a chaotic 3-body interaction which ends with one object being ejected.
   \item \textit{Hardened chaotic} encounters - where the periapsis of the single passes close to the binary and gas dissipation is highly efficient, leading to a prolonged three-body encounter as gas continually removes energy from the system.
\end{itemize}
We show examples of the trajectories and object separations over time for each encounter type in Figure \ref{fig:encounter_types}.
In the glancing encounter case, the BBH and BH do not pass deeply into each other's accretion discs and have typically high approach velocities, therefore dissipation is not efficient and the single does not become bound. As the periapsis of the encounter is large, the BBH is only mildly perturbed. In Figure \ref{fig:encounter_types}, one can observe the encounter induces an eccentricity change from $e\simeq0.05$ to $e\simeq0.14$ in the BBH. In the hierarchical encounter, the periapsis of the encounter is still large compared to the BBH semi-major axis, but the single is successfully captured through gas drag. Provided the single has sufficient angular momentum that subsequent encounters have periapses greater than the BBH semi-major axis, the single BH remains stable in its orbit about the BBH, forming a quasi-stable hierarchical triple system (see panel 2 in Figure \ref{fig:encounter_types}). 

If the periapsis of the single is approximately equal to the initial BBH semi-major axis, the 3-body dynamics become chaotic. If the gas dissipation is less efficient, then the typical apoapses of the objects (with respect to the 3-body COM) will be larger in accordance with the energy of the system (Eq. \ref{eq:E3body}). This gives a larger probability that one of the BHs will be launched with an apoapsis large enough to escape $r_\mathrm{H,t}$ and become ionised by the SMBH. Upon ionisation, the remaining binary is hardened relative to the penultimate temporary BBH in all our simulations. This is expected as it follows that the energy gained to have $r_\mathrm{p,sin}>r_\mathrm{H,t}$ must have come from an energy exchange with the BBH if the previous orbit of the single had $r_\mathrm{p,sin}<\rht$ and $a_\mathrm{sin}<\rht/2$.

The final scenario, \textit{hardened} chaotic encounters, are akin to the temporary chaotic encounters but with more efficient gas drag during the initial interaction and in the following evolution. In these simulations, energy is removed rapidly from the 3-body system, quickly shrinking the mean separation of each of the three BHs, creating a far more compact 3-body system that remains chaotic. As the 3-body system loses energy, the required energy gain of one of the BHs to remove it from the system becomes increasingly large, requiring extremely close encounters (and correspondingly massive binary hardening) to eject a BH. Since this mechanism requires efficient drag to shrink the 3-body system quickly (so that an object is not ejected in the early stages of the 3-body interaction where $a_\mathrm{sin}$ is typically larger) this scenario becomes more common for higher ambient gas densities, as we will see in Sec. \ref{sec:param_space}. 

As a simple test, we switch off gas in our hierarchical example at $t=16\Omega_0^{-1}$ (Figure \ref{fig:encounter_types}, top row, second column). When gas is switched off, we observe the evolution of $a_\mathrm{sin}$ and $a_\mathrm{bin}$ immediately stagnates. Modulations still exist in the separation of the binary + single separation due to accelerations from the SMBH.

\subsection{Hierarchical triples, in more detail}
\begin{figure}
    \centering
    \includegraphics[width=6.7cm]{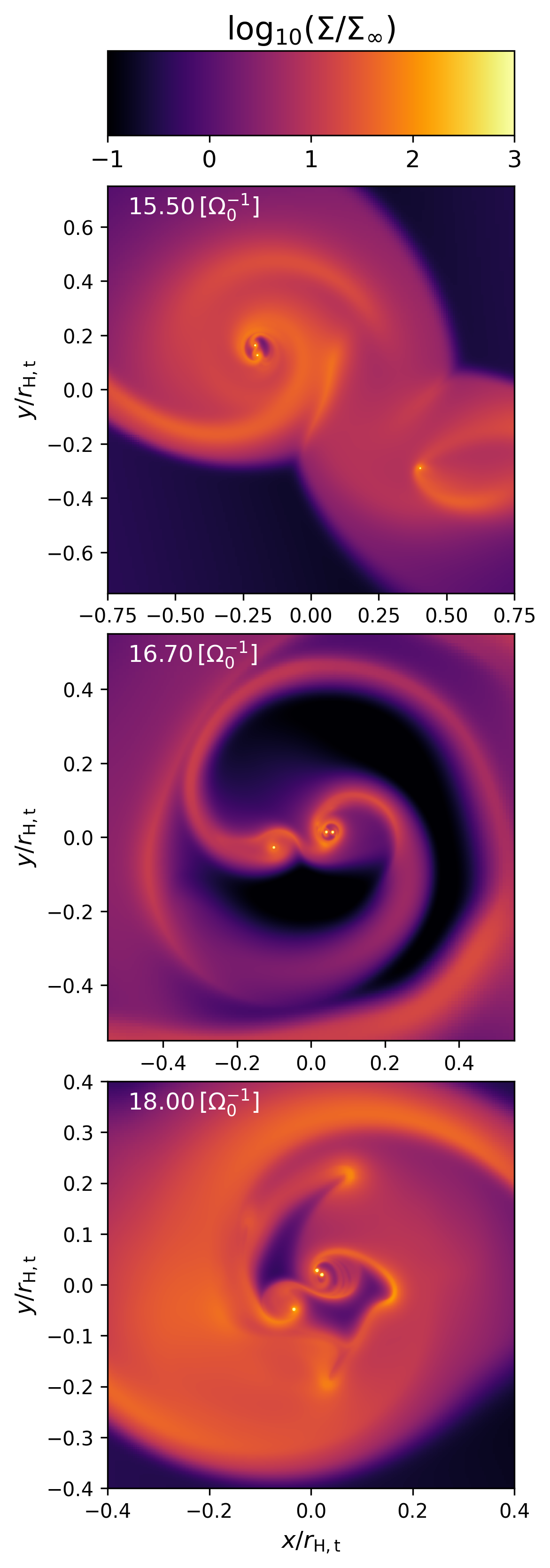}
    \caption{The gas morphology during the formation of and evolution of an embedded hierarchical triple system ($\Sigma=\Sigma_0$, $p=1.8\rht$). \textit{Top:} The intersection of the circum-binary and circum-single discs just prior to the first close encounter. \textit{Middle:} The production of strong spiral outflows that persist for the first $\sim15-20$ orbits of the single around the binary. \textit{Bottom:} Late stage evolution of the system, where the gas has refiled the 3-body Hill sphere to form a circum-triple disc.}
    \label{fig:hierarchical}
\end{figure}

For encounters that have a typically larger periapsis, the binary may still be treated as a single object during the encounter. In such cases, the BH and gas dynamics are akin to the single-single case. After the initial encounter, the binary and single orbit each other, whilst shedding gas through spiral outflows. The gas dissipation then allows the entire 3-body system to remain bound as a hierarchical triple. We show an example of this scenario in Figure \ref{fig:hierarchical}.
In the example, we have the first approach (top panel) of the single where it is captured into a counter clockwise orbit (prograde with the initial circum-binary disc) around the binary via gaseous dissipation. The resulting orbit of the single is relatively circular ($e_\mathrm{sin}\leq0.1$). The circular orbit and the prograde rotation of the single lead to the generation of large spiral wakes (middle panel). At this point, there is a steady and efficient removal of energy from the system, hardening both $a_\mathrm{sin}$ and $a_\mathrm{bin}$, with $e_\mathrm{sin}$ remaining small. In all hierarchical encounters identified, $a_\mathrm{sin}$ shrinks faster than $a_\mathrm{bin}$. If we consider the BBH and the BBH-BH systems in isolation, this is reminiscent of the $da/dt\propto a$ behaviour expected in binary systems \citep[e.g][]{Ishibashi2020}, since $a_\mathrm{sin}>a_\mathrm{bin}$. This behaviour leads to all hierarchical systems gradually moving back towards chaotic type encounters in their late evolution.

We note that the rotation of the gas around the binary component remains prograde with the binary. The maintenance of a low eccentricity in the effective binary system of the hierarchical binary-single system is consistent with the eccentricity damping in prograde embedded binary simulations \citep[e.g][]{Li_and_lai_2022,Dittman2022,Rowan2022}. After $\sim15-20$ orbits of the single, the gas relaxes and replenishes the 3-body Hill sphere, forming a circum-triple disc. We still identify the presence of a circum-single disc around the single and a circum-binary disc around the binary, with each component of the binary now also having their own miniscule circum-single discs!

We find hierarchical systems are only formed at the beginning of binary-single encounters and no such systems form following an initially chaotic scattering. We suspect this is likely due to a lower local gas mass following the first close encounter and the messy and chaotic gas flows not having time to form the morphology of the hierarchical system shown in Figure \ref{fig:hierarchical} (which follows directly from the interaction of the BBH and BH accretion discs). Therefore the gas is unable to sufficiently circularise the orbit of the single and it undergoes another close 3-body scattering. 
\subsection{Hardened chaotic triples}
\label{sec:hardened_encs}
\begin{figure}
    \centering
    \includegraphics[width=7.8cm]{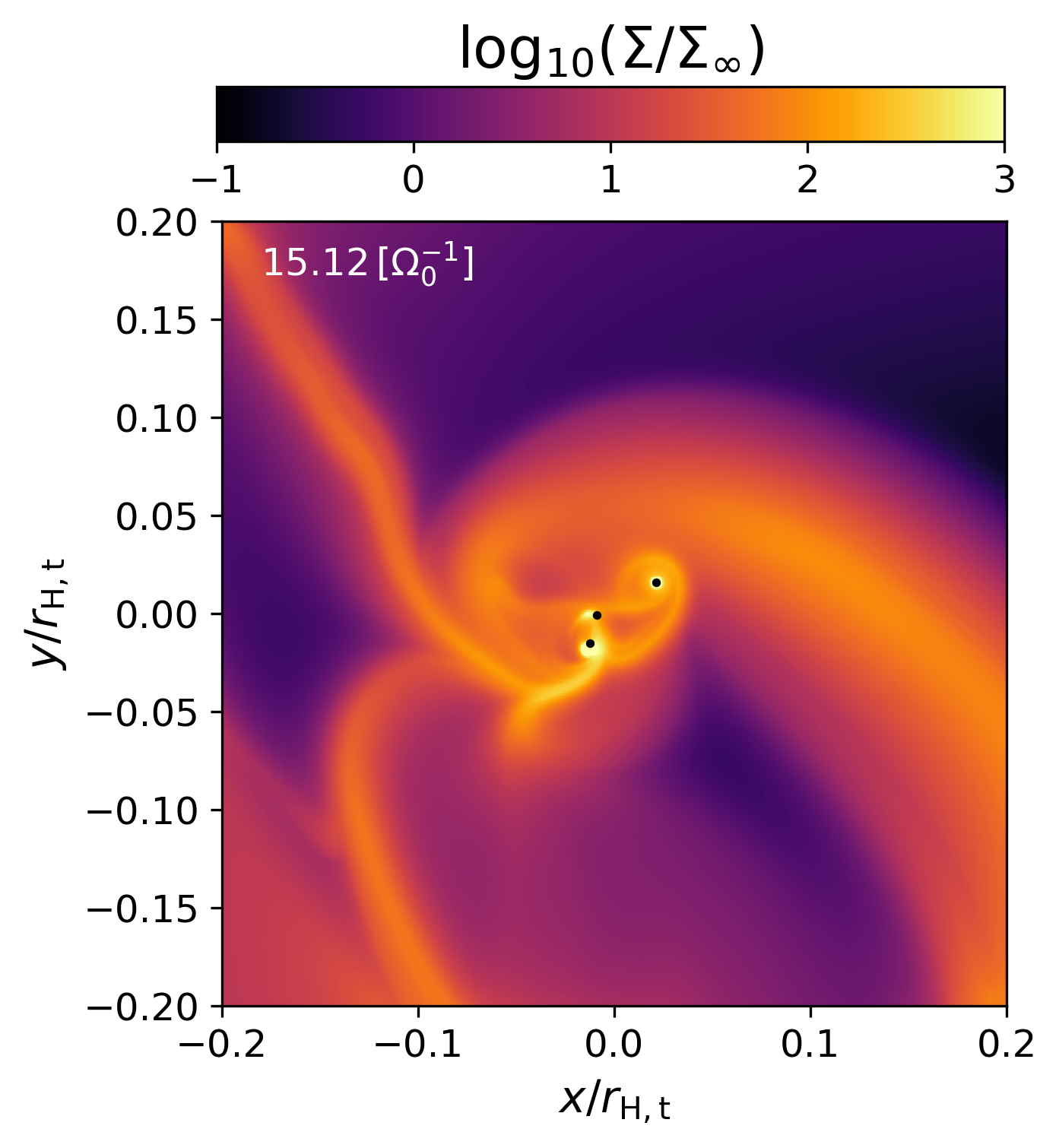}
    \includegraphics[width=7.8cm]{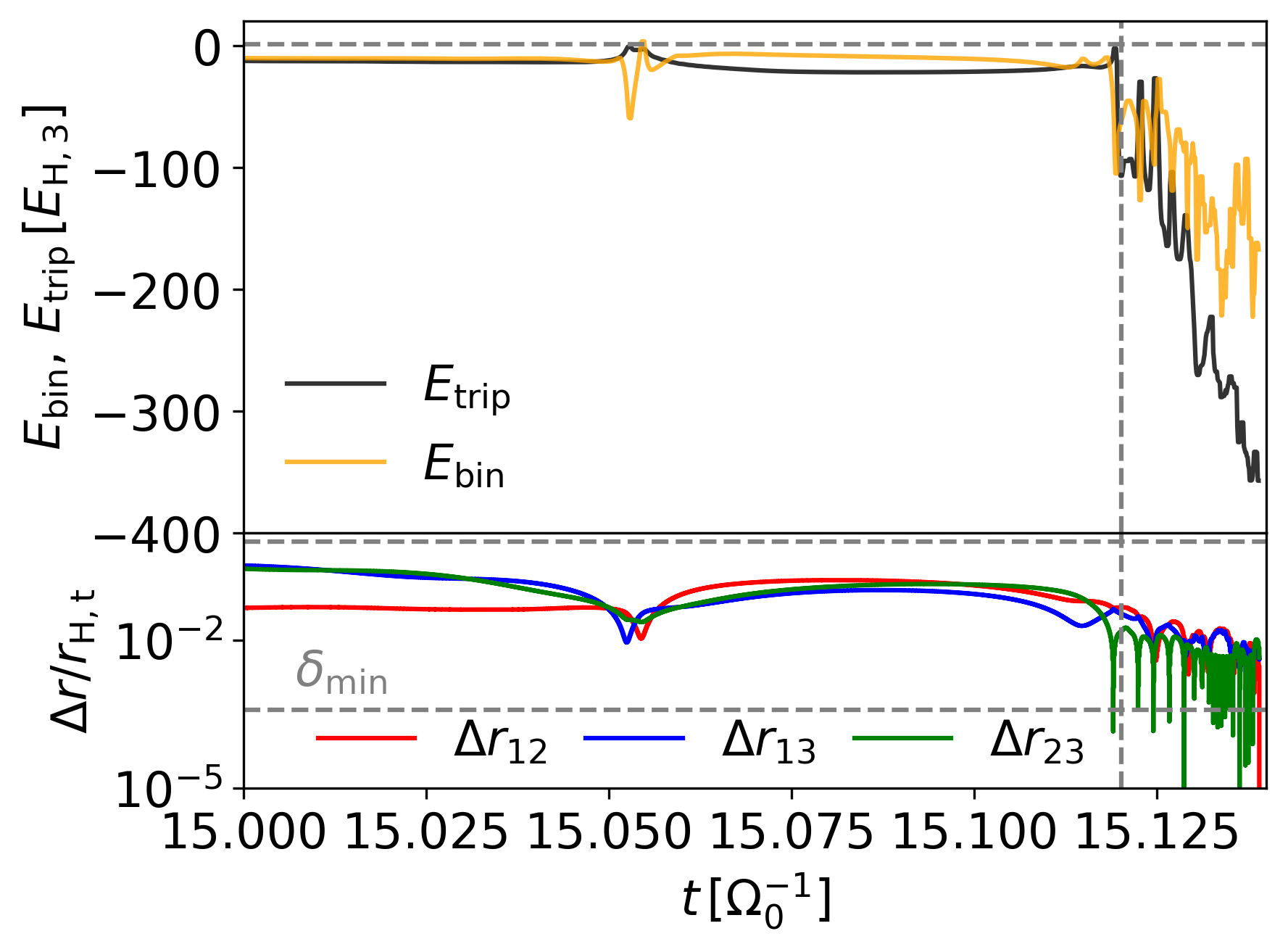}
    \includegraphics[width=8.3cm]{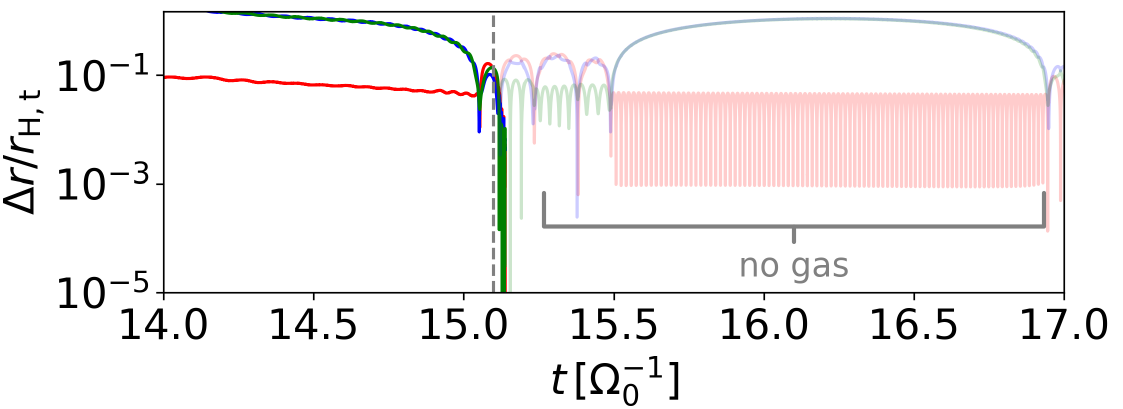}
    \caption{Example of a hardened chaotic encounter, from a run with $p=1.7\rht$ and $\Sigma=\Sigma_0$. \textit{Top:} the gas morphology after the second triple encounter, where the triple becomes strongly bound. Black markers have been added to better indicate the positions of the BHs. \textit{Middle:} the  energy of the triple system $E_\mathrm{trip}$ (Eq. \ref{eq:E3body}), the energy of the most bound BH pair $E_\mathrm{bin}$ (Eq. \ref{eq:E2body}) and BH separations as a function of time. The vertical dashed indicates the time the hydrodynamical snapshot is taken. \textit{Bottom:} The separations between the BHs where gas is turned is off at $t=15.1\Omega_0^{-1}$ (dashed vertical line). The faded lines indicate the triple evolution when gas gravity is switched off.}
    \label{fig:dissipation_hardened}
\end{figure}

Here, we examine the aforementioned hardened chaotic encounters. When the initial periapsis of the single is of the same order as the binary semi-major axis, gas dissipation is highly efficient, forming a tightly bound 3-body system that evolves in a chaotic manner. These systems are arguably the closest hydrodynamical analogue to the gasless 3-body simulations of e.g. \cite{Samsing+2022,Fabj2024,Trani2024}, where the separations between each BH are constantly changing. We show how the 3-body energy and BH separations evolve as an example in Figure \ref{fig:dissipation_hardened}. 
In this example, there is a small initial amount of dissipation following a more shallow first encounter. This softens the initial binary such that the system is no longer a hierarchical system. This is followed by a strong 3-body encounter that efficiently removes energy from the system, reducing the mean separation between \textit{all} BHs. Energy is then efficiently and continuously removed from the system. We observe fluctuations where energy is injected rather than removed from the triple or temporary softenings of the most bound BH pair. However, in all hardened encounters there is a net removal of energy. As another test, we switch off the gas just prior to the first strong encounter (bottom panel of Figure \ref{fig:dissipation_hardened}). When gas is switched off, the system retains a large mean separation between the three objects, i.e. $a_\mathrm{bin}$ and $a_\mathrm{sin}$ remain large during each binary-single state. The absence of gas also leads little change in $a_\mathrm{bin}$ as the single executes its larger orbit, where we would before expect gradual inspiral from gas drag. 

What separates the temporary from hardened encounters is how quickly the system can contract relative to the number orbits of the single. At each close 3-body encounter, the re-shuffling of $a_\mathrm{sin}$ presents another opportunity for the single BH to escape. If the system contracts quickly, then the region of the parameter space for the single ($a_\mathrm{sin}$ and $e_\mathrm{sin}$) that will facilitate its escape quickly becomes smaller. Since the dissipation scales with the density $\Sigma$, we find hardened chaotic encounters are more common in our higher density simulation suites.

\subsection{The parameter space of binary-single encounters}
\label{sec:param_space}
\begin{figure}
    \centering
    \includegraphics[width=8cm]{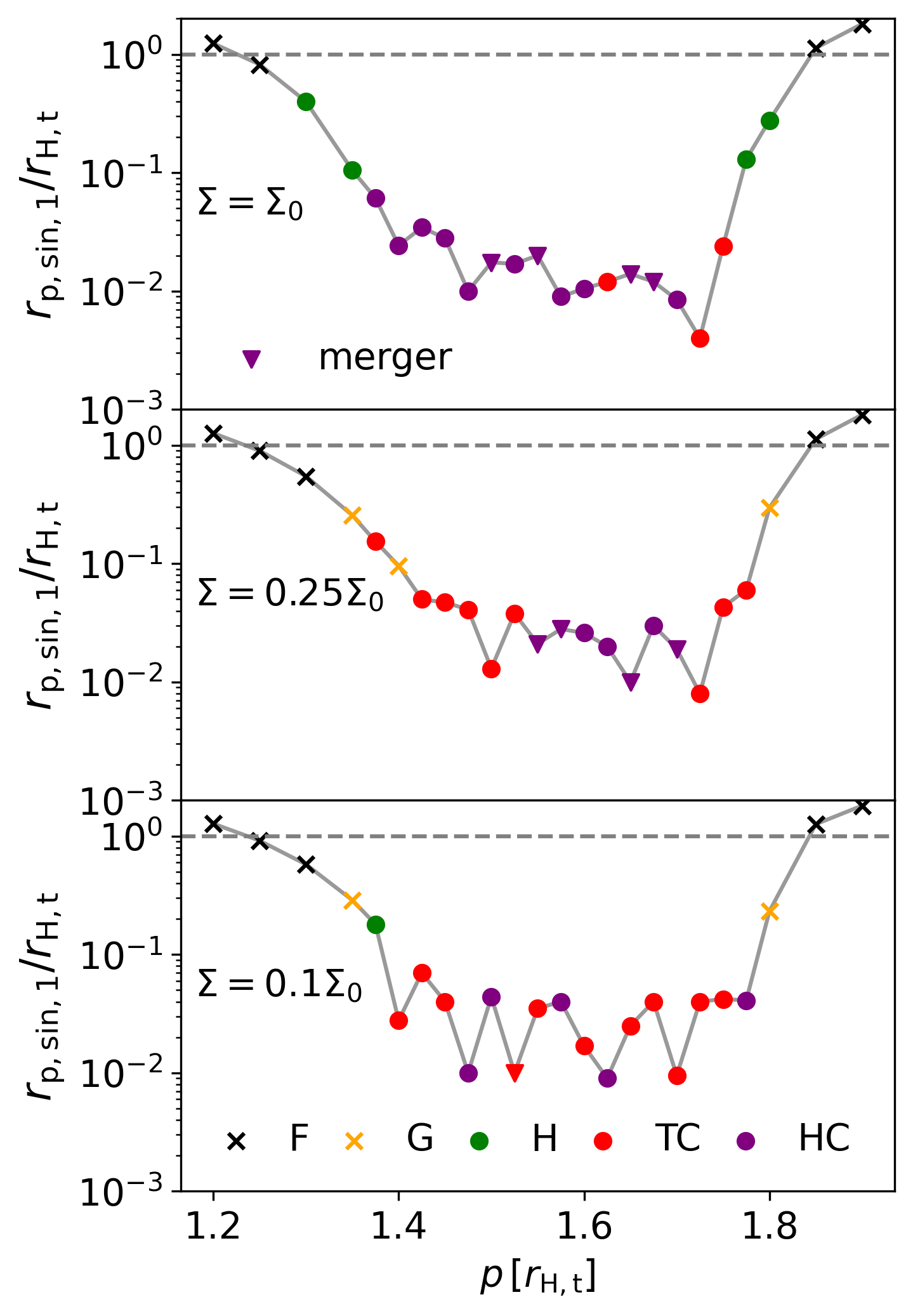}
    \caption{The closest approach of the single $r_\mathrm{p,sin,1}$ as measured from the binary COM as a function of impact parameter $p$ for each simulation suite in density $\Sigma$. The data points are colour coded by the type of encounter:  no close interaction (F), glancing encounter (G), hierarchical encounter (H), temporary chaotic encounter (TC) and hardened chaotic encounter (HC). The results indicate increased AGN disc densities produce more HC encounters. The crossed markers represent triple interactions where there is no second encounter. The triangular markers indicate runs were the separation of two BHs is small enough to reasonably expect a merger while the system was resolved (see Sec. \ref{sec:mergers_secular}).}
    \label{fig:encounter_window}
\end{figure}

To better understand the dependence of the encounter type on the initial encounter parameters, we show the minimum separation of the single and the binary COM as a function of the single's initial impact parameter $p$ and colour code the data points according to the encounter type in Figure \ref{fig:encounter_window}.
The ``w'' shape of the curve is reminiscent of the periapsis-impact parameter distribution of single-single BH scatterings \citep[e.g][]{Boekholt_2022,Whitehead2023,Rowan2023}. The periapsis of the first encounter, $r_\mathrm{p,sin,1}$, in the centre of our impact parameter $p$ domain is more stochastic in the binary single case. This is due to the BBH not being another point mass, so the separation between the BBH COM and the single becomes hard to predict when the single reaches a separation of $a_\mathrm{bin}$ from the BBH. As the original binary hardens slightly less during the approach of the single for lower $\Sigma$, this stochasticity is larger.

Similar to the single-single case, the rotation of the single about the 3-body COM is prograde (counter clockwise) for low impact parameters left of the first trough and higher parameters beyond the second trough. For mergers in-between, the single executes a retrograde trajectory, where the transition between prograde and retrograde encounters occurs on the troughs, representing the cases where the angular momentum of the single as it enters $\rht$ tends towards zero at periapsis. 

As mentioned, hierarchical and glancing encounters occur in prograde encounters with larger initial periapses. For systems in the central retrograde portion of the $p-r_\mathrm{p,sin,1}$ domain, we observe either temporary or hardened chaotic encounters. Comparing the $\Sigma=\Sigma_0$ to the $\Sigma=0.1\Sigma_0$ runs, we find an increase larger than $200\%$ in the number of hardened chaotic encounters (13 vs 5) despite a comparable number of at least temporary formed triples. (20 vs 17), owed to typically more efficient energy dissipation. We summarise the scattering outcomes for each value of the density $\Sigma$ suite in Table \ref{tab:outcomes}.
\begin{table}
\begin{tabular}{|c|c|c|c|c|c|} 

 \hline
     $\Sigma/\Sigma_0$       & F & G & H & TC & HC \\
 \hline
 1  & 4 & 0 & 4 & 3 & 13 \\ 
 \hline
 0.25 & 5  & 3 & 0 & 9 & 6 \\
 \hline
 0.1 & 5  & 2 & 1 & 11 & 5  \\
 \hline
\end{tabular}
\centering
\caption{The number of encounter types per simulation suite with density $\Sigma$. The encounter types are labelled as: failed encounters(F), i.e no close encounter, glancing encounters (G), hierarchical (H), temporary chaotic encounters (TC) and hardened chaotic encounters (HC).}
\label{tab:outcomes}
\end{table}
\subsection{Gravitational wave prospects}
\label{sec:mergers}

\subsubsection{Constraining the merging timescale at the resolution limit}
\label{sec:mergers_secular}
Following our findings that gas reliably hardens the three-body system, we consider the prospects for gravitational wave sources originating from embedded binary-single scatterings. Given the hardened chaotic encounters manage to reach the resolution limit of our simulations, we can infer the fate of the system by anticipating the least favourable outcome for merger. Here, we make pessimistic assumptions for the evolution of the system to give a lower bound on the merger probability. 

\begin{figure}
    \centering
    \includegraphics[width=8.5cm]{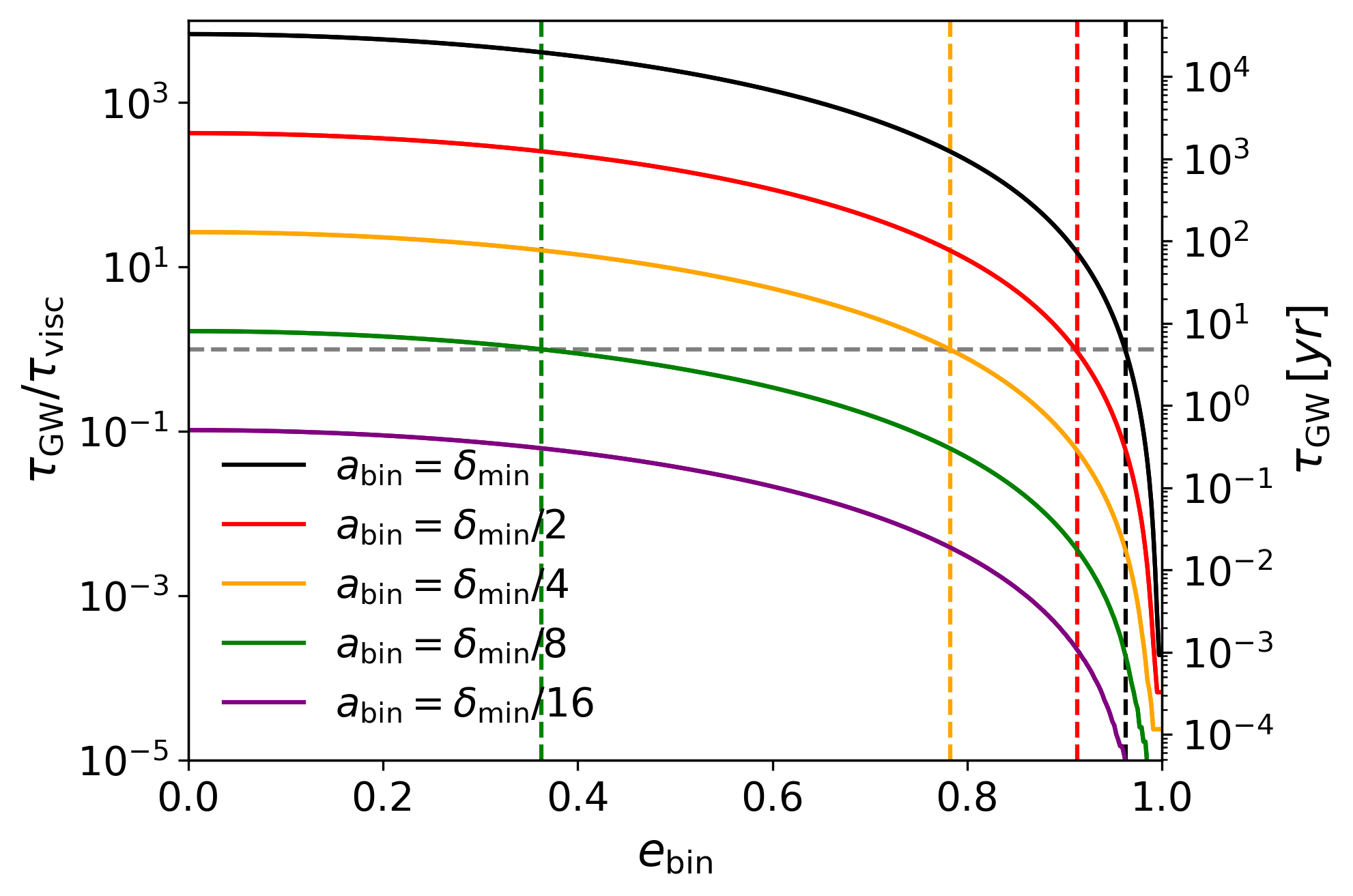}
    \caption{The maximal timescale until a BBH merger, $t_\mathrm{GW}$ for varying initial binary separations around the resolution limit of our simulations $a_\mathrm{bin}=\{1,1/2,1/4,1/8,1/16\}\delta_\mathrm{min}$. These values assume a three-body (or binary) system is not further hardened by gas, which may significantly reduce these values. The timescale is shown in the units of the viscous timescale for the circum-binary disc $\tau_\mathrm{visc}=R_\mathrm{disc}^{2}/\nu$. Whether the binary will merge within the viscous time of the disc depends sensitively on $e_\mathrm{bin}$ and $a_\mathrm{bin}$.}
    \label{fig:merger_timescale}
\end{figure}

If we assume that the 3-body system will continue to efficiently harden at scales smaller than the resolution $\delta_\mathrm{min}$, then the ejection of the third object is the only means to increase the timescale for a BH merger. This follows as the evolution of embedded binaries with separations on the scale of $\delta_\mathrm{min}$ has not yet been simulated self consistently from the larger length scales shown here, therefore we remain agnostic to their hydrodynamical evolution. The ejection of an object also removes the chance of the chaotic 3-body system to result in a very close flyby of two BHs that may induce a merger (see Sec. \ref{sec:mergers_dynamic}). We estimate the hardening of the remaining binary upon the ejection of a BH by considering the energy exchange required to soften $a_\mathrm{sin}=\delta_\mathrm{min}$ to $a_\mathrm{sin}=\rht$. The required gain in energy to escape to a distance $r_2$ from some initial value $r_1$ is given by.
\begin{equation}
    \centering
    E_\mathrm{esc}=\frac{G M_3 (M_1+M_2)}{2}\bigg(\frac{1}{r_1}-\frac{1}{r_2}\bigg)\,,
    \label{eq:E_esc}
\end{equation}
Taking the energy of the binary as $E_\mathrm{bin}=-G M_1 M_2/2a_\mathrm{bin}$, the hardening of the binary is then related to the energy exchange by 
\begin{equation}
    \centering
    \frac{\Delta a_\mathrm{bin}}{a_\mathrm{bin}}=\frac{-E_\mathrm{esc}}{E_\mathrm{bin}}=\frac{M_3(M_1+M_2)}{M_1M_2}\bigg(\frac{a_\mathrm{bin}}{r_2}-\frac{a_\mathrm{bin}}{r_1}\bigg)\,.
    \nonumber
\end{equation}
Setting $r_1=\delta_\mathrm{min}$, $r_2=\rht$ and noting here $M_1=M_2=M_3$, this simplifies to
\begin{equation}
    \centering
    \frac{\Delta a_\mathrm{sin}}{a_\mathrm{sin}}=2\bigg(\frac{a_\mathrm{bin}}{\rht}-\frac{a_\mathrm{bin}}{\delta_\mathrm{min}}\bigg)\,.
    \label{eq:hardening}
\end{equation}
Taking $a_\mathrm{bin}=a_\mathrm{sin}/2=\delta_\mathrm{min}/2$, (i.e the binary is still very soft with respect to the triple system) this gives 
\begin{equation}
    \centering
    \frac{\Delta a_\mathrm{sin}}{a_\mathrm{sin}}=2\bigg(\frac{\delta_\mathrm{min}}{2\rht}-\frac{1}{2}\bigg)\approx-1\,\,\,\,(\delta_\mathrm{min}<<\rht)\,.
    \label{eq:hardening}
\end{equation}
This means the resulting binary  will have a final semi-major axis of $\delta_\mathrm{min}/4\approx0.0019$au (0.4$R_\odot$) under these pessimistic assumptions. We refer to \cite{Fabj2024} for a more detailed explanation of this analytic approximation. Assuming all binaries have this maximal semi-major axis, we compute the GW inspiral time as a function of the eccentricity $e_\mathrm{bin}$ by numerically integrating the orbital evolution equations of \cite{Peters1964}
\begin{equation}
    \centering
    \bigg\langle\frac{da}{dt}\bigg\rangle=-\frac{64}{5}\frac{G^3 M_1 M_2 (M_1 + M_2)}{c^5 a^3(1-e^2)^{7/2}}\bigg(1+\frac{73}{24}e^2+\frac{37}{96}e^4\bigg)\,,
    \label{eq:dadt}
\end{equation}
\begin{equation}
    \centering
    \bigg\langle\frac{de}{dt}\bigg\rangle=-\frac{304}{15}\frac{G^3 M_1 M_2 (M_1 + M_2)}{c^5 a^4(1-e^2)^{5/2}}\bigg(e+\frac{121}{304}e^3\bigg)\,.
    \label{eq:dedt}
\end{equation}
We compare the inspiral timescale $t_\mathrm{GW}$ with the viscous timescale for gas at the outer edge of a circum-binary disc of size\footnote{Once the single is ejected, the binary Hill sphere becomes the appropriate length scale for the formation of a disc.} $R_\mathrm{disc} = \rhb$/2, $\tau_\mathrm{visc}=R_\mathrm{disc}^2/\nu$ as a measure of time to re-establish a typical circum-binary gas morphology. The relative timescales are shown in Figure \ref{fig:merger_timescale} for final binaries with $a_\mathrm{bin}=\{1,1/2,1/4,1/8,1/16\}\delta_\mathrm{min}$. The results show that the length scales of our simulations come close to those where GWs could begin to dominate the evolution of the binary. We can be more confident in the fate of more eccentric binaries, owed to the stronger GW emission at periapsis. From the $1/a^3$ dependence of Eq. \eqref{eq:dadt}, just a factor 4 decrease in separation of $a_\mathrm{bin}=\delta_\mathrm{min}/16$ from the fiducial $a_\mathrm{bin}=\delta_\mathrm{min}/4$ value leads to reliable mergers independently of $e$. Assuming the eccentricity of the final binary follows the theoretical co-planar derived distribution of $P(e_\mathrm{bin})=e_\mathrm{bin}/\sqrt{1-e_\mathrm{bin}^2}$ \cite[e.g][]{Monaghan1976}, the percentage of binaries that meet the criterion $\tau_\mathrm{GW}<\tau_\mathrm{visc}$ is \{28\%, 38\%, 61\%, 92\%, 100\%\} for $a_\mathrm{bin}=\{1, 1/2,1/4,1/8,1/16\}\delta_\mathrm{min}$, implying a non-negligible amount of systems may merge. We stress that this conclusion contains several pessimistic assumptions:
\begin{itemize}
    \item Most notably, for this calculation we assume gas does not further harden the triple system before a BH is ejected, which may still be significant below the resolvable scales in our simulations. 
    
    \item We assume gas does not further harden the resulting binary upon the ejection of the single. If the binary experiences net energy removal at these separations, this would imply all such systems would merge on a timescale shorter than predicted on the right hand axis of Figure \ref{fig:merger_timescale} regardless of whether $\tau_\mathrm{GW}<\tau_\mathrm{visc}$ or not. 

    \item We assume a BH is ejected with the minimum amount of energy to reach $\rht$. In reality the ejected BH can carry more energy away from the system, leaving behind a tighter binary. Where the remaining energy of the binary approximately follows the distribution P($|E_\mathrm{bin}|)\propto |E_\mathrm{bin}|^{-3}$, see \cite{Valtonen2006,Stone2019}.

    \item As we do not modify the viscosity $\nu$ beyond the assumptions of a steady AGN disc, the viscous time will be underestimated owed to the scale height $H=c_\text{s,0}/\Omega=c_\text{s,0}\sqrt{\frac{R^{3}}{GM_\text{bin}}}$
    in the circum-binary disc being smaller than that of the AGN disc (maintaining the isothermal assumption). 

    \item We ignore the prospect for GW dissipation to halt the removal of the single during its final periapsis, which would similarly harden the triple system and potentially move the system into the regime where any resulting binary falls within the regime required for reliable merger.
\end{itemize}
Note that the hydrodynamical evolution of the triple will become inaccurate before reaching the resolution $\delta_\mathrm{min}$. If we take the softening length of our sinks to be the limiting scale of the hydrodynamics\footnote{Given the encounters are typically highly eccentric, the majority of the orbit will be spent at higher separations.} for a three body interaction, then this corresponds to $\sim10\delta_\mathrm{min}$. Therefore we must also trust that the system will harden beyond this point. We note that we find no triple systems are softened by the gas over the range $a=a_0$ to $a=10\delta_\mathrm{min}$, therefore we expect this to be a fair assumption.
\subsubsection{Highly eccentric dynamical mergers}
\label{sec:mergers_dynamic}
In several simulations (9 of 65 initially bound triple systems formed) we find BH separations of $\Delta r_{ij}<10r_\bullet$ where $r_\bullet=GM_\mathrm{BH}/c^{2}$ is the Schwarzschild radius of one of the BHs. These cases are the result of the inherently unstable 3-body configuration, where each encounter randomises the energy and angular momentum of the objects. While we are limited in our simulations to a resolution limit of $\delta_\mathrm{min}\simeq3\times10^{4}r_\bullet$, these cases often occur while the separations of the BHs were thus far fully resolved. As we don't expect the trajectories of the BHs to significantly change in the very small timeframe of the approach from $\Delta r=\delta_\mathrm{min}$ to $\Delta r<10r_\bullet$, we expect the close periapsis to result in strong GW emission and possible prompt merger of the BHs involved. Following a similar methodology to \cite{Fabj2024}, we estimate the probability enhancement for a binary merger with observable residual eccentricity due to random ultra-close encounters during hardened chaotic interactions. 

\begin{figure}
    \centering
    \includegraphics[width=8
cm]{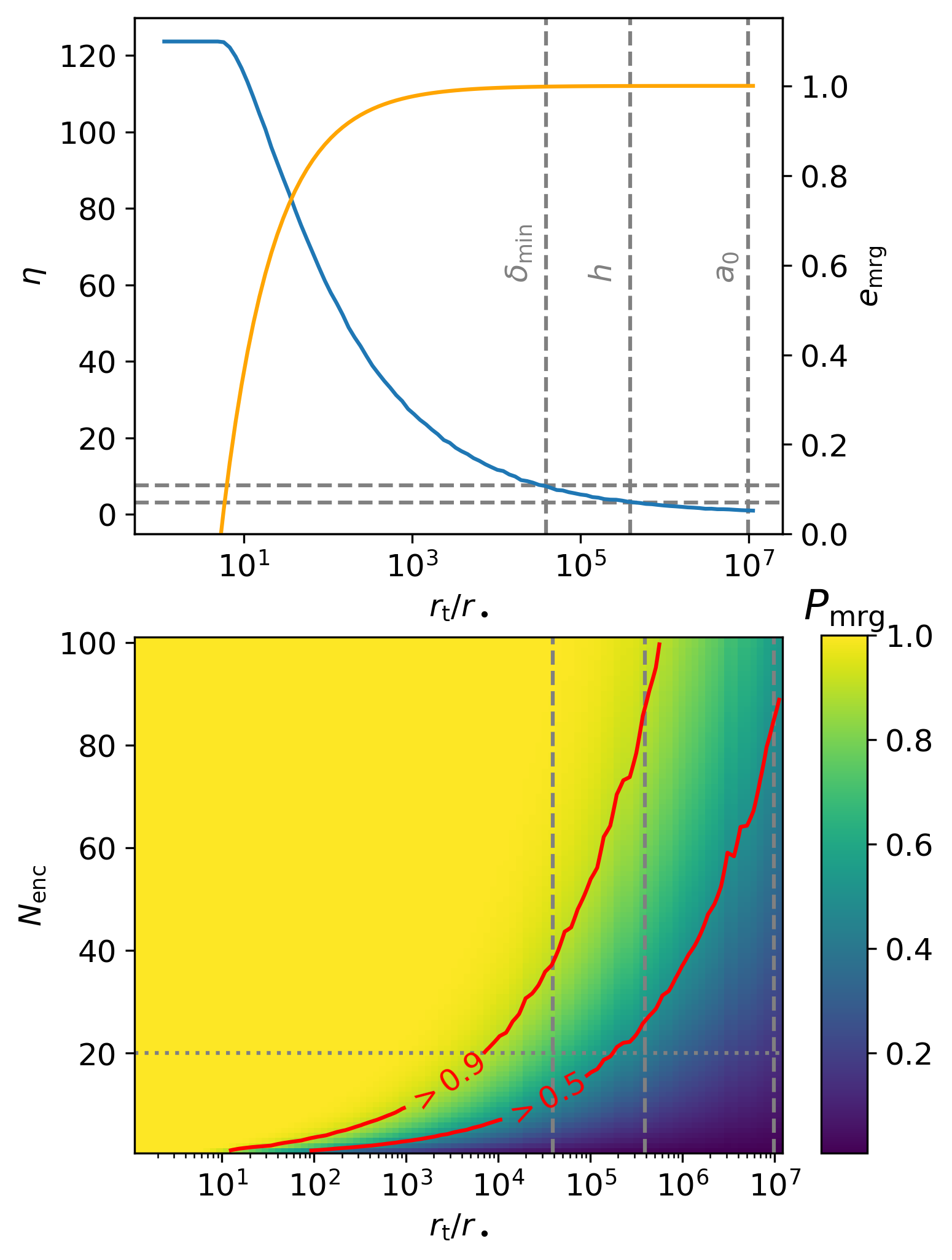}
    \caption{The enhancement of the merger probability due to gas hardening. \textit{Top:} the minimum critical eccentricity $e_\mathrm{mrg}$ (Eq. \ref{eq:e_merge}) required for two BHs to merge during a triple exchange as a function of the compactness of the triple $r_\mathrm{t}$ in Schwarzschild radii. Also shown is the probability enhancement per encounter $\eta$ (Eq. \ref{eq:eta}) that this eccentricity is achieved relative to the initial binary compactness. \textit{Bottom:} The probability of a merger $P_\mathrm{mrg}$ as a function of the number of binary-single like states created during a 3-body exchange $N_\mathrm{enc}$ as a function of the compactness of the triple. We mark the contours of $P_\mathrm{mrg}=\{0.5,0.9\}$. In both plots we indicate the length scales of the resolution $\delta_\mathrm{min}$, the softening $h$ and initial binary semi-major axis $a_0$.} 
    \label{fig:eccentric_mergers}
\end{figure}

The inspiral time of a BBH in the high eccentricity limit is given by 
\begin{equation}
    t_\mathrm{GW,e}=\frac{5\sqrt{2}c^{5}}{64G^{3}}\frac{\abin^{1/2}\pbin^{7/2}}{M_1 M_2 (M_1+M_2)}\,.
    \centering
    \label{eq:t_gw_e}
\end{equation}
Assuming the current 3-body dynamics are un-hierarchical and the triple system has a characteristic scale $r_\mathrm{t}$ (i.e $\abin \sim \asin \sim r_\mathrm{t}$), then the orbital period of the single around the binary is on the order of 
\begin{equation}
    T_\mathrm{sin}\approx2\pi\sqrt{\frac{r_\mathrm{t}^{3}}{G(M_1+M_2+M_3)}}\,.
    \centering
    \label{eq:t_gw_e}
\end{equation}
For the binary to merge before the single as the chance to change the binary dynamics, we must have $t_\mathrm{GW,e}<T_\mathrm{sin}$. Again assuming only the equal mass case $M=M_1=M_2=M_3$ this gives a maximal periapsis for the binary to reliably merge of
\begin{equation}
    r_\mathrm{mrg}(r_\mathrm{t}) \lesssim \bigg(\frac{256\pi G^{5/2}}{5\sqrt{6}c^{5}}\bigg)^{2/7}r_\mathrm{t}^{2/7}M^{5/7}\,,
    \centering
    \label{eq:rp_merge}
\end{equation}
where we note a small factor $3^{-2/7}$ correction to equation (3) in \cite{Fabj2024}.
Expressing this as an eccentricity $e_\mathrm{mrg} = 1-r_\mathrm{mrg}/\abin$, we have 
\begin{equation}
    e_\mathrm{mrg}(r_\mathrm{t}) \gtrsim 1 - \bigg(\frac{256\pi G^{5/2}}{5\sqrt{6}c^{5}}\bigg)^{2/7}r_\mathrm{t}^{-5/7}M^{5/7}\,.
    \centering
    \label{eq:e_merge}
\end{equation}
In the absence of gas, the scale of the triple system is set by the initial binary separation $a_0$, as the system cannot contract since there is no dissipation\footnote{In principle chance encounters that are very close could dissipate energy via GWs but not induce merger, but we assume this to be overall far less efficient than gas on these larger scales.}. Therefore the minimal\footnote{The value is a minimum as it assumes the number of encounters is unaffected by gas, whereas in reality gas will induce a drag on objects that are on initially escape trajectories, increasing $N_\mathrm{enc}$. We do not have enough statistics here to comment on this directly.} enhancement $\eta$ to the probability of a merger due to gas is the ratio of the probability that $e>e_\mathrm{mrg}(r_\mathrm{t})$, i.e.
\begin{equation}
    \eta=\frac{P(e>e_\mathrm{mrg}(r_\mathrm{t}))}{P(e>e_\mathrm{mrg}(a_0))}\,.
    \centering
    \label{eq:eta}
\end{equation}
Recalling $P(e)\approx e/\sqrt{1-e^2}$ in the co-planar limit, we compute these probabilities as a function of $r_\mathrm{t}$ in Figure \ref{fig:eccentric_mergers}. We also show the probability that a 3-body system will undergo a merger as a function of the number of encounters and $r_\mathrm{t}$, assuming the probability scales as $P_\mathrm{mrg}(r_\mathrm{t})=1-[1-P(e>e_\mathrm{mrg}(r_\mathrm{t}))]^{N_\mathrm{enc}}$, where $N_\mathrm{enc}$ is the number of encounters.
For triples hardened to $r_\mathrm{t}=h$ and $r_\mathrm{t}=\delta_\mathrm{min}$ through gaseous drag, we find the probability per encounter to have $e>e_\mathrm{crit}$ is enhanced by a factor $\eta\approx 3.5$ and $8$, respectively, compared to $r_\mathrm{t}=\rht$. Adopting the mean number of encounters of $N_\mathrm{enc}\approx20$ as suggested by \cite{Fabj2024}, we find the probability for systems that harden to $r_\mathrm{t}=h$ and $\delta_\mathrm{min}$ to merge is $\sim0.4$ and $\sim 0.65$. We note that the addition of gas often leads to longer 3-body exchanges in addition to prolonged periods where the identification of a binary-single like state is not possible, although whether the latter result enhances or reduces the likelihood of a highly eccentric encounter between two BHs is uncertain.
\section{Discussion and Implications}
\label{sec:discussion}
Our results suggest that the combination of gas drag and the complex 3-body dynamics of binary-single scatterings in AGN discs could lead to the rapid merger of previously secularly evolving embedded BBHs. This would indicate that predictions for the number of required binary-single encounters to harden a system to merger from N-body or analytic studies that do not include gas are an upper bound. For example, \cite{Leigh2018} predict a mean of $\sim$10 such encounters to dynamically harden a binary enough to inspiral via GWs. The degree of this overestimation is sensitive to the density of the AGN disc, therefore we stress the need for constraints on this quantity, which can vary wildly depending on which AGN disc model and/or viscosity is assumed e.g. Thompson \citep{Thompson_disc_2005} or Sirko-Goodman discs \citep{Sirko2003,Goodman2004} and their physical parameters (e.g SMBH accretion rate).

Though we only consider co-planar binary-single encounters with no velocity dispersions, the 3-body simulations of \cite{Trani2024} around an SMBH find that the distribution of final eccentricities and semi-major axes are largely unaffected provided the dispersion $\sigma$ is $\sigma 
<10^{-4}v_\mathrm{K}$, where $v_\mathrm{K}$ is the Keplerian orbital velocity about the SMBH. Therefore we do not expect the 3-body dynamics to differ significantly within this threshold. At higher dispersions, they predict the remaining binary eccentricities will follow a more thermal rather than super-thermal distribution, reducing the merger likelihood. The scale height $H/R \approx0.0056\sim 10^{-2}$ provides the length scale where the gas density is approximately unchanged and where we expect only small changes in the magnitude of gaseous triple hardening. However, we suspect there will be a stronger inclination dependence on the level of dissipation during the first encounter depending on the relative inclination of the encounter to the circum-single/binary discs, which warrants further study. 

The reliable shrinkage of the triple system that we report has several implications not considered in the main body of this work. The reduction in the mean value of $a_\mathrm{bin}$ and more importantly $a_\mathrm{sin}$ increases the likelihood that a third object may be in close proximity to a merging binary during the exchange. This implies that triples in AGN could produce unique dephasing features in the GW waveform of the inspiralling binary \citep[e.g][]{Samsing2024,Hendriks2024_2,Hendriks2024}. By a similar argument, a tighter 3-body system increases the chance that a tight binary with a small enough separation and large enough eccentricity to have a measurable residual eccentricity. Such cases could potentially be detected as more sensitive third-generation detectors come online, with Cosmic Explorer \citep[e.g][]{Reitze2019,Evans2023,Borhanian2024} and the Einstein Telescope \citep[e.g][]{Hild2008,Franciolini2023} design sensitivities able to detect $e\gtrsim5\times10^{-3}$ and $e\gtrsim10^{-3}$, respectively at a GW frequency of $10$Hz \citep[e.g][]{Saini2024}. The hardening of embedded triples also increases the binding energy between a merging binary and the tertiary. This increases the prospect for the now 2-body system to be retained after the remnant receives a post-merger kick \citep[e.g][]{Gonzales2007,Campanelli2007,Varma2022}, allowing for possible back to back or ``double'' mergers \citep[e.g][]{Samsing2019}. 

In principle, the gas drag during a chaotic encounter of a binary and a single BH directly relates to the scenario of binary-binary scatterings \citep[e.g][]{Mikkola1984,Bacon1996,Heggie2000,Arca_Sedda+2021}, which serve to reduce the binary fraction on average \citep[e.g][]{Sweatman2007,Sollima2008,Ryu2017}. There is no reason to suggest that gas drag will not dissipate energy during the initial encounter of two binaries and during the chaotic exchange that follows, hardening the system. We therefore posit that possible reductions in the merger rate due to binary-binary (and binary-single) encounters in an AGN disc will be mitigated to some degree when gas is included, since the merger probability will be enhanced. Similarly to the hierarchical triple scenario, it is plausible that gas could aid in the formation of highly exotic hierarchical (2+2) 4-body systems, provided the impact parameter is relatively large compared to the binaries' semi-major axes.

\section{Summary and Conclusions}
\label{sec:conclusions}
We performed 72 hydrodynamical simulations of binary-single black hole interactions in an AGN disc covering 3 disc densities and 24 impact parameters, examining the role of gas on the fate of the triple system.  We summarise our key findings below:
\begin{itemize}
    \item Similar to single-single BH encounters, binary-single interactions feature significant energy dissipation during the first encounter, allowing for the formation of an energetically bound triple system.
    \item We identify four characteristic interactions for close binary-single encounters in gas (Sec. \ref{sec:encounter_types}): \textit{Glancing encounters}, where the single does not become bound but modifies the eccentricity and/or semi-major axis of the original binary. \textit{Hierarchical encounters}, where the single is bound through gas dissipation into a low eccentricity orbit around the original binary, forming a quasi-stable hierarchical triple system. \textit{Temporary chaotic encounters}, where the periapsis of the single is on par with the initial binary separation, leading to a chaotic encounter that ends with one object being thrown out. \textit{Hardened chaotic encounters}, where the dissipation during a chaotic encounter is efficient enough to quickly reduce the mean separation between the three BHs, leading to prolonged three-body encounters. 
    \item For the chaotic encounters (Sec. \ref{sec:hardened_encs}), the gas morphology is highly dynamic and volatile, leading to fluctuations in the energy of the 3-body system. Although, in all cases, net energy is removed from the triple.
    \item Assuming the system undergoes a close encounter with separations less than the 3-body Hill sphere, chaotic encounters are the most common encounter type for our range of AGN disc densities $\Sigma$ (Sec. \ref{sec:param_space}).
    \item The type of encounter is primarily set by first periapsis distance of the single object $r_\mathrm{p,sin,1}$ and the gas density $\Sigma$. Glancing and hierarchical encounters occur at larger $r_\mathrm{p,sin,1}$ and $\Sigma$, temporary chaotic encounters are more likely at small  $r_\mathrm{p,sin,1}$ and $\Sigma$. Chaotic hardened encounters occur at small $r_\mathrm{p,sin,1}$ and large $\Sigma$ (Figure \ref{fig:encounter_window}).
    \item Provided the triple remains bound, gas continually removes energy from the three-body system, causing it to contract. For hardened chaotic encounters, this shrinks the typical semi-major axes of the BBH and BBH-BH components during the establishment of binary-single like states during the three-body interaction. For hierarchical encounters, we find the semi-major axis of the BBH-BH system shrinks faster than the BBH, gradually bringing the system to a more chaotic state. The efficiency of the dissipation in all cases scales with the density of the AGN disc. 
    \item Dissipation is most efficient in hardened chaotic encounters, where the 3-body system shrinks to our resolution limit $\delta_\mathrm{min}$ in a few AGN orbits, a length scale two orders of magnitude smaller than the initial binary separation. We evolved the orbital elements of binaries at the resolution limit, assuming a BH is immediately ejected (Sec. \ref{sec:mergers_dynamic}). Sampling over the eccentricity distribution, we find 28\% of our chaotic systems are expected to merge through secular GW emission. 
    \item Depending on the compactness $r_\mathrm{t}$ of the triple, the critical minimum eccentricity that two BHs must have in order to merge during a binary + single state is given in Eq. \eqref{eq:e_merge}. The relation predicts a $\sim$3.5 and 8 times increase in the probability (during a single encounter) for the three body system to merge after being hardened to the softening $h$ and resolution $\delta_\mathrm{min}$ lengths respectively (Sec. \ref{sec:mergers_secular}). For a fiducial value of 20 binary + single states, the probability of merger is $\sim0.4$ and $\sim0.65$ for $r_\mathrm{t}=h$ and $r_\mathrm{t}=\delta$, respectively (Sec. \ref{sec:mergers_dynamic}). This suggests that while they are potentially rare, binary-single encounters in AGN discs could reliably induce BH mergers.
    \item The ability for gas to shrink the size of the 3-body system will increase the likelihood of unique BH-BH merging scenarios. This includes potential observable eccentricity in GWs, dephasing from the tertiary object and repeated mergers. As we are limited by our resolution limit we leave the quantitative analysis of these sub-grid phenomena to future research.
\end{itemize}
We have made several assumptions and simplifications in this work. We have restricted ourselves to co-planar binary-single encounters. In practice there will be some vertical and radial velocity dispersion, which may alter the outcomes of the initial scattering and subsequent evolution \citep[e.g][]{Trani2024}. We neglect the accretion of mass by the stellar BHs which may be 
significant \citep[][]{Li_and_Lai_2022_windtunnel_II,Rowan2022}, although we previously found that softening lengths of comparable radii to the accretion boundary lead to similar dissipation \citep[e.g][]{Rowan2023}. We only consider equal mass BHs here, whereas the true range may lie in the region of $\sim5\mbox{--} \,40\,\msun$ (depending on the assumed metalicity of the nuclear stellar cluster, e.g \citealt[][]{Belczynski2010b}). For computational reasons, we assumed an isothermal equation of state. Our previous simulations of single-single scatterings utilising a non-isothermal equation of state (\citealt{Whitehead2023_novae}, see also the work of \citealt[][]{Li_and_Lai_2022_windtunnel_II}) reported non-negligible mass reduction in the Hill sphere of each BH before and during the first encounter. Since the dissipation scales with the local density, we expect that evolving the fluid energy equation will produce results more akin to our lower density runs (i.e more temporary than hardened chaotic encounters compared to the isothermal case).


AGN discs host a large range of dynamical and hydrodynamical problems. As in single-single scatterings, we conclude that gas drag is an important component of binary-single encounters and the evolution of embedded 3-body systems, noticeably increasing the likelihood of a merger during their evolution. Current approaches to incorporate the effects of gas dynamical drag in few-body systems are usually based on employing gas drag formulas which are valid only for rectilinear \citep[e.g][]{Ostriker1999} or circular \citep[e.g][]{Kim2007,Kim2008} trajectories of the massive bodies even when more complex motions are involved. Recent studies hold the promise to consider the effects of gas under more general conditions \citep[e.g][]{ONeill2024,Suzuguchi}. Our findings here, combined with our previous work on single-single scatterings \citep{Rowan2022,Rowan2023,Whitehead2023,Whitehead2023_novae}, provide motivation and guidance for the future development of more accurate analytical, or semi-analytical, models to capture gas effects on single-single and binary-single black hole evolution and mergers in AGN. These effective models could then be used in population studies to produce more accurate estimations for the GW observables arising from black hole mergers in AGN discs accounting explicitly for the effects of gas.
\section*{Acknowledgements}
\begin{itemize}

\item  This work was supported by the Science and Technology Facilities Council Grant Number ST/W000903/1 attributed to Bence Kocsis. 

\item The research leading to this work was supported by the Independent Research Fund Denmark via grant ID 10.46540/3103-00205B attributed to Martin Pessah. 

\item This was work was supported by the Villum Fonden grant No. 29466, and by the
ERC Starting Grant no. 101043143 — BlackHoleMergs attributed to Johan Samsing.

\item This work was performed using resources provided by the Cambridge Service for Data Driven Discovery (CSD3) operated by the University of Cambridge Research Computing Service (www.csd3.cam.ac.uk), provided by Dell EMC and Intel using Tier-2 funding from the Engineering and Physical Sciences Research Council (capital grant EP/T022159/1), and DiRAC funding from the Science and Technology Facilities Council (www.dirac.ac.uk).

\end{itemize}
\bibliographystyle{mnras}
\bibliography{Paper} 
\label{lastpage}
\end{document}